\documentclass[10pt,twocolumn]{IEEEtran}

\usepackage[T1]{fontenc}
\usepackage[latin9]{inputenc}

\usepackage{amsmath,amssymb,amsbsy,amsthm,amstext}
\usepackage{mathrsfs}
\usepackage{bbm}
\usepackage{comment}
\usepackage{graphicx}
\usepackage{array}
\usepackage{tabularx}
\usepackage{multirow}
\usepackage{makecell}
\usepackage{cellspace}
\usepackage[table,x11names,dvipsnames]{xcolor}

\usepackage{algorithm}
\usepackage{algorithmic}

\usepackage{float}
\usepackage{cite}
\usepackage{eurosym}

\usepackage{hyperref,soul}

\definecolor{mygray}{gray}{0.6}
\definecolor{myblue}{rgb}{0.8,0.85,1}

\newcolumntype{L}[1]{>{\raggedright\arraybackslash}p{#1}}
\newcolumntype{C}[1]{>{\centering\arraybackslash}m{#1}}
\newcolumntype{R}[1]{>{\raggedleft\arraybackslash}m{#1}}
\newcolumntype{Y}{>{\raggedright\arraybackslash}X}
\newcommand{\tcell}[2]{%
\parbox[t]{#1}{\raggedright #2}%
}
\DeclareRobustCommand{\officialeuro}{%
  \ifmmode\expandafter\text\fi
  {\fontencoding{U}\fontfamily{eurosym}\selectfont e}}

\begin{document}

\title{\huge Artificial Intelligence for Spatially Reconfigurable Antennas: Movable, Fluid, and Pinching Antenna Systems}

\author{Nguyen Cong Luong, Zeping Sui, \textit{Member, IEEE}, Thai-Hoc Vu, Jie Cao, Bo Ma, Thuan Van Le, Xunyang Zhan, Nguyen Duc Hai, Min Xu, Qiushi Zhao, Dong In Kim,~\IEEEmembership{Life Fellow,~IEEE}, Yonghong Zeng,~\IEEEmembership{Fellow,~IEEE}, and Shaohan Feng, \textit{Member, IEEE}
\vspace{-2em}

\thanks{Nguyen Cong Luong and Nguyen Duc Hai are with the Phenikaa School of Computing, Phenikaa University, Hanoi 12116, Vietnam (e-mails: luong.nguyencong@phenikaa-uni.edu.vn, hai.nguyenduc@phenikaa-uni.edu.vn).}

\thanks{Zeping Sui is with the School of Computer Science and Electronics Engineering, University of Essex, Colchester CO4 3SQ, U.K. (e-mail: zepingsui@outlook.com).}

\thanks{Thai-Hoc Vu is with the Faculty of Electrical Engineering and Computer Science, VSB-Technical University of Ostrava, 17. Listopadu 2172/15, 708 00, Ostrava, Czechia (e-mail: thai.hoc.vu@vsb.cz).}

\thanks{Jie Cao and Xunyang Zhan are with the Faculty of Information and Communication Engineering, Harbin Institute of Technology, Shenzhen (e-mails: caojhitsz@ieee.org, 24s052039@stu.hit.edu.cn).}

\thanks{Bo Ma, Qiushi Zhao, and Shaohan Feng are with the School of Information and Electronic Engineering (Sussex Artificial Intelligence Institute), Zhejiang Gongshang University, Hangzhou 310018,
China. (e-mails: 25020090093@pop.zjgsu.edu.cn, \{mabo, feng\_shaohan\}@mail.zjgsu.edu.cn).}

\thanks{Thuan Van Le is with the Faculty of Electrical and Electronic Engineering, Phenikaa School of Engineering, Phenikaa University, Hanoi 12116, Vietnam (e-mail: thuan.levan@phenikaa-uni.edu.vn).}

\thanks{Min Xu is with the School of Mathematics, Statistics and Mechanics, Beijing University of Technology, Beijing 100124, China (e-mail: xm@bjut.edu.cn).}

\thanks{Dong In Kim is with the Department of Electrical and Computer Engineering, Sungkyunkwan University, Suwon 16419, South Korea (e-mail: dongin@skku.edu).}

\thanks{Yonghong Zeng is with the I2R, A*STAR, Singapore 138632 (email: yhzeng@ieee.org).}
}

\maketitle

%====================================================================
\begin{abstract}
Recently, sixth-generation (6G) wireless networks have moved beyond fixed-array designs toward antenna architectures that can adapt their spatial configuration to specific environmental conditions. Movable antenna, fluid antenna, and pinching antenna systems represent this principle in different ways, but they share a common vision: exploiting spatial flexibility as an additional degree of freedom (DoF) to improve communication, sensing, security, and resource efficiency. These new techniques, however, also bring challenging problems, as antenna configuration must be jointly considered with channel acquisition, beamforming, mobility, and network resource management. Therefore, artificial intelligence (AI) has become an important tool for learning fast and adaptive control policies for these highly coupled systems. In this survey, we provide a unified review of AI for spatially reconfigurable antenna systems. We first introduce the basic principles of movable, fluid, and pinching antennas, which is followed by a summary of the latest AI-enabled designs according to their primary optimization objectives. Furthermore, we compare the roles of deep learning (DL), deep reinforcement learning (DRL), multi-agent reinforcement learning (MARL), graph learning, Transformers, large language models (LLMs), and structure-guided learning across different antenna architectures. Finally, we discuss open challenges and future directions toward scalable, robust, and hardware-aware intelligent reconfigurable antenna networks. 
\end{abstract}


%=====================================================================
\section{Introduction}
%=====================================================================

% \IEEEPARstart{T}{he} ever-increasing demand for higher spectral efficiency, lower latency, and more reliable connectivity in 6G wireless networks
% has pushed the boundaries of conventional antenna design. Traditional
% fixed-position antenna (FPA) systems are fundamentally constrained by their
% static spatial configuration: once deployed, the positions of antenna elements cannot be adjusted to respond to changing channel conditions or user distributions. This rigidity limits the system's ability to fully exploit the spatial degrees of freedom (DoF) available in the propagation environment, leaving significant performance gains unrealized.

\IEEEPARstart{F}{uture} wireless networks are expected to support increasingly dense, mobile, and service-diverse environments \cite{sui2025multi,chirp1,rou2026afdm}. This evolution marks the antenna system as no longer merely a passive interface between a transceiver and the propagation medium. Instead, it becomes an active part of wireless system design. Traditional fixed-position antenna (FPA) systems are fundamentally constrained by their static spatial configuration: once deployed, the positions of antenna elements cannot be adjusted to respond to
the time-varying conditions of channels, user distributions, or sensing targets \cite{wang2025robust,wang2025ambiguity}. This rigidity constrains the system's ability to exploit the spatial DoF available in the propagation environment fully, leaving significant performance gains unrealized. Thus, a new line of antenna technologies is urgently needed, in which the spatial configuration itself becomes controllable. 

A new class of \textit{reconfigurable antenna} technologies has recently emerged as a promising answer to this limitation. Among them, \textit{movable antennas} (MAs), \textit{fluid antenna systems} (FASs), and \textit {pinching antenna systems} (PASSs) have attracted significant research interest. MAs allow antenna elements to be repositioned within a bounded region, so that the transceiver can exploit local spatial channel variations and improve beamforming performance \cite{wang2026learning, liang2026two, weng2024learning, shao2025hybrid}. Meanwhile, FASs achieve spatial flexibility via a compact set of candidate ports, with the active port selected based on the instantaneous channel state or interference conditions \cite{new2024tutorial, wang2024ai}. PASSs utilize flexible waveguides and pinching elements to generate radiating points along a deployed structure, enabling the system to place radiation sources closer to users and reshape the propagation geometry more directly \cite{xu2026joint, xie2025graph}. Although these technologies differ in their physical realization, they are unified by a common objective: exploiting spatial reconfigurability as a new optimization dimension in wireless communication systems, whether in antenna positions, ports, or radiating locations. 

This additional spatial DoF creates new opportunities in communication, sensing, security, energy efficiency, and edge intelligence. For communication, spatial reconfiguration can improve channel gain, reduce inter-user interference, and provide additional flexibility for multiple access. For sensing and integrated sensing and communication (ISAC), it can reshape the effective aperture and improve spatial resolution. For physical-layer security and anti-jamming, it can strengthen the legitimate channel while suppressing undesired links. For energy- and latency-sensitive systems, it can reduce the burden on transmit power and resource allocation by improving the underlying channel before digital optimization. These gains explain why MA, FAS, and PASS have been studied rapidly across diverse scenarios, including multi-user multiple-input multiple-output (MIMO), cell-free (CF) networks \cite{sui2024ris,sui2024star,sui2025performance}, non-terrestrial networks, uncrewed aerial vehicle (UAV) communications, integrated sensing and communications (ISAC), simultaneous wireless information and power transfer (SWIPT), federated learning, and mobile edge computing.

However, spatial reconfigurability also changes the nature of wireless system optimization. The antenna configuration is coupled with numerous factors, including channel acquisition, beamforming, user scheduling, power control, trajectory design, phase-shift optimization, and task-oriented resource allocation. In MA systems, this coupling often appears through continuous antenna-position variables and position-dependent channel responses. In FASs, it appears through discrete port selection and interactions between ports and users. In PASSs, it appears through waveguide-constrained radiating positions and near-field propagation effects. Therefore, many resulting problems are high-dimensional, non-convex, mixed discrete-continuous, and time-varying. Classical optimization remains essential for modeling, benchmarking, and deriving structural insights, but repeated online optimization can be costly when channels, users, and service demands change rapidly. 

Therefore, AI has become an important tool for spatially reconfigurable antenna systems, as it can learn quickly and adaptively map across wireless environments, antenna configurations, and resource allocation decisions. For MAs, AI addresses the continuous coupling between antenna positions and channel responses: learning-based methods infer antenna locations and beamforming vectors from channel or location features, while deep unfolding and two-timescale learning embed optimization structures to reduce online complexity and separate slow antenna movement from fast beamforming updates \cite{weng2024learning, wang2026learning, liang2026two}. For FASs, AI is often utilized to handle large port selection spaces, sparse port observations, and switching decisions: reinforcement learning (RL) supports fluid antenna multiple access (FAMA) and opportunistic FAMA by avoiding exhaustive port search \cite{waqar2024opportunistic}, attention-based and self-supervised models extrapolate full port-domain channel state information (CSI) from limited observations \cite{zhang2024learning,zhao2026taaformer}, and graph neural networks (GNNs) or LLMs have recently been explored for joint port selection and beamforming in large combinatorial spaces \cite{guo2025llm,he2025graph}. For PASSs, AI can capture the waveguide-constrained, near-field nature of the system, in which pinching antenna positions jointly affect path loss, phase, beamforming, and sensing performance. Data-driven and graph-based models provide low-latency pinching antenna placement and beamforming inference \cite{kang2025campass, guo2025graph}, while structure-guided learning embeds Karush--Kuhn--Tucker (KKT) conditions to preserve optimization structure, as well as improve generalization beyond black-box prediction \cite{xu2026joint}. Recent DL estimators, DRL controllers, and graph-enhanced LLM frameworks further show the potential of AI for PASS channel estimation (CE), resource allocation, and ISAC \cite{xiao2025channel,lv2026deep,gao2026llm}. These developments suggest that AI is not merely an additional technique, but a common design framework for making spatial reconfigurability practical.

% Deep learning can provide fast mappings from channel or location features to antenna configurations and beamforming variables. DRL and MARL are useful when antenna control is sequential, dynamic, or distributed. Graph neural networks (GNNs) can exploit the relational structure among users, ports, antennas, and waveguides. Transformers and LLMs have recently been explored for channel prediction, port selection, and heuristic search in large combinatorial spaces \cite{guo2025llm,wang2025llm,he2025graph}. Structure-guided learning, such as deep unfolding and neural designs guided by Karush-Kahn-Tucker (KKT) conditions, further shows that learning can benefit from optimization theory rather than replacing it entirely \cite{liang2025two,xu2026joint}. 

Several surveys and tutorials have discussed related topics. Regarding MA systems, the authors in \cite{zhu2023movable} provided an overview of promising applications of MA-aided wireless communications, while the tutorial in \cite{zhu2025tutorial} presented new field-response channel models tailored for MAs, applicable to narrowband and wideband systems, as well as to far-field and near-field propagation conditions. For FASs, the tutorial in \cite{new2024tutorial} provided a comprehensive treatment of FASs from communication theory, optimization, and hardware perspectives. Meanwhile, the authors in \cite{wang2024ai} focused on AI-empowered FAS, discussing challenges and future research directions. Regarding PASSs, the work in \cite{liu2025pinching} reviewed the features of PASSs relative to conventional wireless systems, analyzed their benefits, and discussed potential designs. In contrast, the authors in \cite{yang2025pinching} focused on different scenarios involving PASSs with different numbers of waveguides and antennas, as well as promising 6G-related applications of PASSs.  More broadly, the work in \cite{zhao2026reconfigurable} surveyed reconfigurable antenna technologies for next-generation networks. However, none of these works provided a comprehensive review of AI-empowered techniques across all three reconfigurable antenna paradigms. In parallel, many recent works have discussed the utilization of AI in wireless network systems, including advanced DL for 6G \cite{jiao2024advanced}, agentic GNNs for wireless communications \cite{lu2026agentic}, AI reasoning for networking \cite{luo2025ai}, and Generative AI for next-generation communications \cite{khoramnejad2025generative}. Nevertheless, none of these works surveyed how AI is integrated into spatially reconfigurable antenna frameworks. A unified AI-oriented view across MA, FAS, and PASS is still missing.

% many recent works have proposed learning-based methods for individual spatial reconfigurable systems, including MA-assisted beamforming and channel estimation \cite{weng2024learning}, \cite{liang2026two}, \cite{feng2026deep}, FAS port selection and port-domain channel extrapolation \cite{waqar2024opportunistic}, \cite{zhang2024learning}, \cite{zhao2026taaformer}, as well as PASS beamforming, channel estimation, and ISAC design \cite{xie2025graph}, \cite{xu2026joint}, \cite{gao2026llm}. 

This survey fills a gap in the literature by reviewing AI for spatially reconfigurable antenna systems from a joint-optimization perspective. Instead of organizing the literature only by deployment scenario, we focus on the main coupling that each work addresses, such as antenna configuration with beamforming, channel estimation, sensing, trajectory or phase control, security, and resource allocation. This taxonomy makes it easier to compare different antenna architectures under the same design objective, and to understand which AI tools are suitable for which type of coupling. It also highlights the physical differences among MA, FAS, and PASS: MA emphasizes continuous local movement, FAS emphasizes fast port-domain selection, and PASS emphasizes waveguide-enabled near-user radiation. The main contributions of this survey are summarized as follows:
\begin{itemize}
\item We provide a unified tutorial on MA, FAS, and PASS, covering operating principles, channel models, and key physical differences. We also explain why their non-convex, high-dimensional, mixed discrete-continuous, and time-varying design motivates the use of AI.
\item We review AI-enabled MA designs for antenna positioning, beamforming, channel estimation, ISAC, trajectory and phase-shift control, security, and resource allocation, including deterministic policy-gradient RL, deep unfolding, meta-learning, and Transformers.

\item We survey AI-enabled FAS designs for port selection, user admission, beamforming, channel estimation, sensing, trajectory control, resource allocation, and offloading, focusing on RL, GNNs, and LLMs.

\item We review AI-enabled PASS designs for PA placement, beamforming, resource allocation, and sensing, and discuss how data-driven, KKT-guided, and graph-based methods exploit their near-field and waveguide-constrained structure.

\item We compare the three architectures from a learning perspective. Particularly, we compare the roles of different AI models, including DL, DRL, MARL, GNNs, Transformers, LLMs, deep unfolding, and structure-guided learning, and explain how their suitability depends on the underlying antenna physics.

\item Finally, we discuss key challenges and future research directions, including CSI acquisition for training, scalability, hardware and sim-to-real issues, AI overhead, and generalization. We also highlight promising directions in cell-free and near-field learning, integration with other 6G technologies, hybrid architectures, energy efficiency, security, and standardization.
\end{itemize}

\iffalse

\begin{itemize}
    \item First, we provide a unified introduction to MA, FAS, and PASS, clarifying their operating principles, channel characteristics, and physical distinctions.
    \item We review recent AI-enabled designs for each architecture in terms of joint optimization problems, covering beamforming, channel estimation, ISAC, trajectory and phase-shift design, security, and resource allocation.
    \item We compare the roles of different AI models, including DL, DRL, MARL, GNNs, Transformers, LLMs, deep unfolding, and structure-guided learning, and explain how their suitability depends on the underlying antenna physics.
    \item Finally, we discuss key challenges and future research directions, including CSI dependence, coupled optimization, scalability, hardware and sim-to-real issues, robustness, AI overhead, and generalization. We also highlight promising directions in cell-free and near-field learning, integration with other 6G technologies, hybrid architectures, energy efficiency, security, and standardization.
\end{itemize}
\fi

The remainder of this survey is organized as follows. Section \ref{sec:fundamentals} introduces the fundamentals of MA, FAS, and PASS. Sections \ref{sec:ai_for_MAs}, \ref{sec:ai_for_FAs}, and \ref{sec:ai_for_PASS} survey
techniques applied to MA, FAS, and PASS systems, respectively, organized by
joint optimization problem. Finally, Section \ref{sec:conclusions} discusses open issues and future research directions and concludes the survey. The major acronyms used in this paper are summarized in Table~\ref{tab:acronyms}. Method-specific abbreviations appearing in the summary tables follow the names used in the corresponding references.

\begin{table}[t]
\centering
\footnotesize
\caption{List of acronyms frequently used in this paper}
\label{tab:acronyms}
\renewcommand{\arraystretch}{1}
\begin{tabular}{|l|l|}
\hline
\rowcolor{myblue}\textbf{Acronym} & \textbf{Definition} \\
\hline
A2C & Advantage actor-critic \\
CNN & Convolutional neural network \\
CSI & Channel state information \\
CTDE & Centralized training with decentralized execution \\
DDPG & Deep deterministic policy gradient \\
DDQN & Double deep Q-network \\
DNN & Deep neural network \\
DoF & Degree of freedom \\
DQN & Deep Q-network \\
DRL & Deep reinforcement learning \\
FAMA & Fluid antenna multiple access \\
FAS & Fluid antenna system \\
FPA & Fixed-position antenna \\
GNN & Graph neural network \\
ISAC & Integrated sensing and communication \\
LLM & Large language model \\
MA & Movable antenna \\
MADDPG & Multi-agent deep deterministic policy gradient \\
MARL & Multi-agent reinforcement learning \\
MLP & Multi-layer perceptron \\
NOMA & Non-orthogonal multiple access \\
NTN & Non-terrestrial network \\
OTFS & Orthogonal time frequency space \\
PA & Pinching antenna \\
PPO & Proximal policy optimization \\
RIS & Reconfigurable intelligent surface \\
RL & Reinforcement learning \\
RSMA & Rate-splitting multiple access \\
SAC & Soft actor--critic \\
SIC & Successive interference cancellation \\
SWIPT & Simultaneous wireless information and power transfer \\
TD3 & Twin delayed deep deterministic policy gradient \\
UAV & Unmanned aerial vehicle \\
URLLC & Ultra-reliable low-latency communication \\
\hline
\end{tabular}
\vspace{-2em}
\end{table}

\section{Fundamentals of Movable Antenna, Fluid Antenna, and Pinching Antenna}
\label{sec:fundamentals}
% ============================================================
This section introduces the fundamental principles, system models, and channel
characteristics of MA, FAS, and PASS. Therein, we provide a comparative analysis
of the three technologies, discuss the limitations of conventional optimization
approaches, and explain why AI is the natural and indispensable solution
framework for all three systems. Throughout this section, antenna or radiating-element positions are denoted
by spatial variables, while beamforming vectors are denoted by
\(\mathbf w_k\). For MA, the position variable is a continuous Cartesian
coordinate, for FAS, it is a discrete port index, and for PASS, it is a
continuous longitudinal coordinate along the waveguide.
\subsection{Movable Antenna (MA)}
% TODO: System model, channel characterization, 1D/2D/6DMA variants
\subsubsection{System Model and Operating Principle}

Movable antenna systems differ from conventional MIMO transceivers by
allowing antenna elements to be repositioned within a bounded spatial region,
thereby adapting the physical channel to the communication objective rather
than relying on a fixed spatial geometry~\cite{wang2026learning}. In a typical MA-enabled MIMO link, the transmitter (Tx) is equipped with
\(N_{\rm t}\) MA elements that can be repositioned within a bounded planar
transmit region \(\mathcal{C}_{\rm t}\subset\mathbb{R}^{3}\), subject to a
minimum inter-element spacing \(d_{\min}\) to mitigate mutual coupling.
Similarly, the receiver (Rx) is equipped with \(N_{\rm r}\) MA elements
that can be repositioned within a bounded planar receive region
\(\mathcal{C}_{\rm r}\subset\mathbb{R}^{3}\). The position of the \(n\)-th Tx antenna is denoted by
\(\mathbf{t}_n=[x_{{\rm t},n},y_{{\rm t},n},0]^{\top}\in\mathcal{C}_{\rm t}\),
\(n=1,\ldots,N_{\rm t}\), and the position of the \(q\)-th Rx antenna is
denoted by
\(\mathbf{r}_q=[x_{{\rm r},q},y_{{\rm r},q},0]^{\top}\in\mathcal{C}_{\rm r}\),
\(q=1,\ldots,N_{\rm r}\). The Tx and Rx
position matrices are defined as
\begin{equation}
\mathbf{P}_{\rm t}
=
[\mathbf{t}_1,\ldots,\mathbf{t}_{N_{\rm t}}]
\in\mathbb{R}^{3\times N_{\rm t}},
\
\mathbf{P}_{\rm r}
=
[\mathbf{r}_1,\ldots,\mathbf{r}_{N_{\rm r}}]
\in\mathbb{R}^{3\times N_{\rm r}}.
\label{eq:ma_position_matrices}
\end{equation}

\subsubsection{Channel Model}

The MA channel exploits the spatial non-stationarity of the wireless
environment: channel quality varies across different positions within the
movable regions due to local variations in path gains and phases. The
field-response characterizes the channel model between the Tx and Rx~\cite{liang2026two,weng2024learning}, which captures the
continuous variation of the channel response with respect to the Tx and Rx
antenna positions.

Specifically, for a channel with \(L_{\rm t}\) Tx paths and \(L_{\rm r}\)
Rx paths, let \(\theta_{{\rm t},j},\phi_{{\rm t},j}\) and
\(\theta_{{\rm r},i},\phi_{{\rm r},i}\) denote the elevation and azimuth
angles of departure (AoD) and arrival (AoA) of the \(j\)-th Tx path and
\(i\)-th Rx path, respectively. The corresponding Tx wave vector is given by
\begin{equation}
\mathbf{k}_{{\rm t},j}
=
[\cos\theta_{{\rm t},j}\cos\phi_{{\rm t},j},
\cos\theta_{{\rm t},j}\sin\phi_{{\rm t},j},
\sin\theta_{{\rm t},j}]^{\top}
\in\mathbb{R}^{3\times1},
\end{equation}
and the Rx wave vector \(\mathbf{k}_{{\rm r},i}\in\mathbb{R}^{3\times1}\)
is defined similarly. For a generic Tx antenna position
\(\mathbf{t}\in\mathcal{C}_{\rm t}\), the Tx field response vector (FRV)
is defined as~\cite{weng2024learning}
\begin{equation}
  \mathbf{g}(\mathbf{t}) =
  \left[
    e^{j\frac{2\pi}{\lambda}\mathbf{k}_{{\rm t},1}^{\top}\mathbf{t}},
    e^{j\frac{2\pi}{\lambda}\mathbf{k}_{{\rm t},2}^{\top}\mathbf{t}},
    \ldots,
    e^{j\frac{2\pi}{\lambda}\mathbf{k}_{{\rm t},L_{\rm t}}^{\top}\mathbf{t}}
  \right]^{\top}
  \in\mathbb{C}^{L_{\rm t}\times1}.
  \label{eq:ma_frv_tx}
\end{equation}
Similarly, for a generic Rx antenna position
\(\mathbf{r}\in\mathcal{C}_{\rm r}\), the Rx FRV is
\begin{equation}
  \mathbf{f}(\mathbf{r}) =
  \left[
    e^{j\frac{2\pi}{\lambda}\mathbf{k}_{{\rm r},1}^{\top}\mathbf{r}},
    \ldots,
    e^{j\frac{2\pi}{\lambda}\mathbf{k}_{{\rm r},L_{\rm r}}^{\top}\mathbf{r}}
  \right]^{\top}
  \in\mathbb{C}^{L_{\rm r}\times1}.
  \label{eq:ma_frv_rx}
\end{equation}

We define the path response matrix (PRM)
\(\boldsymbol{\Sigma}\in\mathbb{C}^{L_{\rm r}\times L_{\rm t}}\), whose
\((i,j)\)-th entry represents the response coefficient between the
\(j\)-th Tx path and the \(i\)-th Rx path. The baseband equivalent channel
between a Tx antenna at \(\mathbf{t}\) and an Rx antenna at \(\mathbf{r}\)
is given by~\cite{liang2026two}
\begin{equation}
  h(\mathbf{t},\mathbf{r})
  =
  \mathbf{f}(\mathbf{r})^{\rm H}
  \boldsymbol{\Sigma}
  \mathbf{g}(\mathbf{t}).
  \label{eq:ma_channel}
\end{equation}

For a MIMO link with \(N_{\rm t}\) Tx antennas and \(N_{\rm r}\) Rx
antennas, the Tx and Rx field response matrices (FRMs) are defined as
\begin{align}
\mathbf{G}(\mathbf{P}_{\rm t})
&=
[\mathbf{g}(\mathbf{t}_1),\ldots,
\mathbf{g}(\mathbf{t}_{N_{\rm t}})]
\in\mathbb{C}^{L_{\rm t}\times N_{\rm t}},
\label{eq:ma_tx_frm}\\
\mathbf{F}(\mathbf{P}_{\rm r})
&=
[\mathbf{f}(\mathbf{r}_1),\ldots,
\mathbf{f}(\mathbf{r}_{N_{\rm r}})]
\in\mathbb{C}^{L_{\rm r}\times N_{\rm r}}.
\label{eq:ma_rx_frm}
\end{align}
The resulting MIMO channel matrix is expressed as
\begin{equation}
  \mathbf{H}(\mathbf{P}_{\rm t},\mathbf{P}_{\rm r})
  =
  \mathbf{F}(\mathbf{P}_{\rm r})^{\rm H}
  \boldsymbol{\Sigma}
  \mathbf{G}(\mathbf{P}_{\rm t})
  \in\mathbb{C}^{N_{\rm r}\times N_{\rm t}}.
  \label{eq:ma_mimo}
\end{equation}
The key insight is that repositioning the Tx and Rx antennas changes the
FRVs in~\eqref{eq:ma_frv_tx}--\eqref{eq:ma_frv_rx}, which alters the
constructive or destructive superposition of propagation paths in
\eqref{eq:ma_channel}. Repositioning, therefore, allows the system to seek antenna positions that yield improved channel gain or spatial separability.

In the near-field regime, which becomes relevant for large movable apertures
or millimeter-wave/terahertz frequencies, the spherical wavefront model
replaces the planar wave approximation when the Rayleigh distance
\(2D^2/\lambda\), with physical aperture size \(D\), exceeds the Tx--Rx
separation. We denote \(\mathbf{r}_0\) as the coordinates of the Rx reference
point in the Tx local coordinate system, and \(\mathbf{T}\) as the
coordinate transform matrix between the two systems. The line-of-sight
(LoS) channel component between a Tx antenna at \(\mathbf{t}\) and an Rx
antenna at \(\mathbf{r}\) is~\cite{shao2025hybrid}
\begin{equation}
  h_{\rm LoS}(\mathbf{t},\mathbf{r})
  =
  \sigma_0
  e^{-j\frac{2\pi}{\lambda}
  \|\mathbf{r}_0+\mathbf{T}^{\top}\mathbf{r}-\mathbf{t}\|_2},
  \label{eq:ma_nf_los}
\end{equation}
where \(\sigma_0\) is the LoS path amplitude gain, including
distance-dependent path loss. The non-line-of-sight (NLoS) component
involves point scatterers and can be constructed similarly using
near-field FRVs. The complete near-field channel is
\begin{equation}
  h_{\rm near}(\mathbf{t},\mathbf{r})
  =
  h_{\rm LoS}(\mathbf{t},\mathbf{r})
  +
  h_{\rm NLoS}(\mathbf{t},\mathbf{r}).
  \label{eq:ma_near_channel}
\end{equation}

\subsubsection{MA Variants}

Several MA architectures have been proposed in the literature.
One-dimensional (1D) MA restricts movement to a line segment,
simplifying the optimization while still achieving significant gains over
FPA baselines. Two-dimensional (2D) MA allows movement within a
planar region, offering richer spatial DoF but a larger optimization domain.
Six-dimensional MA (6DMA) further enables control of antenna
orientation and three-dimensional positioning, maximizing spatial DoF for
distributed massive MIMO deployments~\cite{shao2025hybrid}.
Pixel-based MA discretizes the movable region into a grid and
activates subsets of fixed pixels, bridging MA and reconfigurable aperture
concepts~\cite{huang2025deep}. In all variants, the joint optimization of
antenna positions and beamforming is the key research challenge.

\subsubsection{Unique Advantages of MAs}

%Beyond simply providing an additional optimization variable, MA offers a set of physically unique advantages over conventional FPA systems as follows:

\paragraph{Channel Gain Improvement}
MA exploits the position-dependent phases of the multipath components to reach antenna positions with a larger effective channel gain than that at a
prescribed fixed reference location. The achievable gain depends on the
size of the movable region, the path geometry, and the available channel
state information. In particular, MA provides no spatial gain in an
ideal single-path far-field channel, whereas richer multipath propagation
creates more opportunities for constructive channel combining
\cite{zhu2024modeling,weng2024learning}. In multi-user
settings, MA can reduce inter-user channel correlation by adjusting the
array geometry, thereby improving spatial multiplexing and sum-rate
performance \cite{liang2026two,wang2026learning}. A two-timescale framework
separates slow antenna repositioning from fast beamforming adaptation,
achieving near-optimal ergodic sum-rate with low overhead~\cite{liang2025two}.
In the near-field regime, MA additionally exploits both angle and distance
dimensions, yielding a provably higher channel matrix rank than
FPA~\cite{shao2025hybrid,yang2026effective}. 

\paragraph{Interference Management and Multiple Access Enhancement}
By introducing antenna position as an additional DoF, MA can reshape the
effective channel geometry to strengthen desired links and create more
favorable interference-suppression conditions. This provides many benefits to the existing networks. In cognitive radio networks, MA steers a null toward the primary receiver to protect it from interference while maintaining full array gain toward the secondary receiver, which is difficult for FPA to achieve simultaneously. In coordinated multi-point (CoMP) and CF systems, joint position-beamforming optimization suppresses inter-cell interference to improve cell-edge performance~\cite{amhaz2026enhancing}. For non-orthogonal multiple access (NOMA), MA reshapes the channel gain disparity between paired users to strengthen successive interference cancellation (SIC) decoding~\cite{amhaz2025optimizing}.

\paragraph{Energy Efficiency and Wireless Power Transfer}
By adaptively selecting antenna positions with stronger effective channel
gains, MA achieves the same quality of service (QoS) with substantially lower transmit power than
FPA~\cite{amiri2025movable}, which is suitable to green
communications and power-constrained Internet of Things (IoT). In SWIPT systems, MA provides an additional spatial DoF for shaping the
effective information and energy-transfer channels, enabling a more flexible
rate-energy trade-off than FPA-based systems~\cite{amiri2025movable,xie2024movable}.

\paragraph{Sensing Performance}
MA can expand the effective sensing aperture without increasing the number
of antenna elements or radio-frequency (RF) chains. By adjusting the
element locations within the movable region, MA can increase the spatial
spread of the sensing array, which generally improves angular resolution
and can reduce the Cram\'{e}r--Rao bound (CRB) for AoA
estimation under suitable channel and signal-to-noise-ratio conditions
\cite{xiu2026robust,xiu2025movable}. In ISAC systems, the reconfigurable geometry enables a Pareto-optimal capacity--CRB trade-off
inaccessible to fixed-array ISAC~\cite{xiu2026meta,wang2025unsupervised}.
By adjusting the antenna positions, MA can reduce undesired beam replicas
associated with sparse array layouts, thereby improving sensing beampattern
quality and communication beamforming performance~\cite{zhang2026crosstalk,le2025beamforming}.

\paragraph{Physical Layer Security}
MA can enhance the channel gain toward the legitimate receiver while
suppressing the effective channel observed by the eavesdropper, even when
the two receivers are spatially close~\cite{wang2026moving}. Small antenna
displacements create position-dependent phase variations across the
legitimate and wiretap channels, providing an additional spatial DoF for
channel separation beyond conventional fixed-array beamforming. Under
suitable channel assumptions, such positional DoF can help satisfy the
steering-vector orthogonality (SVO) condition, so that the desired channel
is strengthened while the eavesdropper channel is substantially weakened
without excessive beamforming-power loss~\cite{wang2026moving,li2025can}.
For covert communications, MA can reduce the warden's received signal power
toward the noise floor~\cite{xie2024movable}. In anti-jamming scenarios,
MA improves robustness by weakening the effective jamming channel or
suppressing the received jamming power, potentially with fewer antenna
elements than FPA-based designs~\cite{tang2025deep,guo2025movable}.

%A Transformer-based framework can further predict eavesdropper mobility and adapt MA positions proactively~\cite{wang2026moving}.

\iffalse
\paragraph{Channel Estimation and CSI Acquisition}
MA optimization requires channel knowledge across candidate antenna positions,
which can lead to substantially higher CSI acquisition overhead than FPA
systems with fixed array geometry. Recent works reduce this overhead by
reconstructing position-dependent channel maps from a finite number of
measurements, or by learning antenna-position decisions directly from partial
channel-power measurements~\cite{11500467,shao2025hybrid,jang2025new,lu2025learning}.
\fi

\subsection{Fluid Antenna System}
% TODO: Port-based model, spatial diversity, FAMA concept
\subsubsection{System Model and Port-Based Architecture}

In an FAS, a radiating element can physically or electronically select one
among \(N\) predefined ports distributed within a compact aperture of
normalized length \(W_s\) wavelengths~\cite{new2024tutorial}. In its basic single-element form, exactly one port is active at a time,
and the system selects the active port index
\(n_k^\star\in\{1,\ldots,N\}\) for user \(k\). The \(N\) ports are
uniformly distributed over a one-dimensional FAS aperture of normalized
length \(W_s\), where \(\lambda\) denotes the carrier wavelength.
Accordingly, the physical aperture length is \(W_s\lambda\), and the physical inter-port spacing is \(\Delta=W_s\lambda/(N-1)\). In multi-element
FAS (e.g., MIMO-FAS), multiple ports can be simultaneously active across
different antenna branches, each independently selecting its best
port~\cite{new2024tutorial,guo2025llm}.

We consider a multi-Rx FAS downlink in which the Tx is equipped with an
FAS array consisting of \(N_{\rm t}\) radiating elements. Each element can
select one among \(N\) predefined ports within its compact aperture. Let
\(\boldsymbol{\ell}=[\ell_1,\ldots,\ell_{N_{\rm t}}]^{\top}\), with
\(\ell_m\in\{1,\ldots,N\}\), denote the active Tx port configuration, and
let \(\mathbf{h}_k(\boldsymbol{\ell})\in\mathbb{C}^{N_{\rm t}\times1}\)
denote the resulting channel vector from the Tx array to the \(k\)-th
Rx. The received signal at Rx \(k\) is expressed as
\begin{equation}
  y_k(\boldsymbol{\ell})
  =
  \mathbf{h}_k(\boldsymbol{\ell})^{\rm H}\mathbf{w}_k s_k
  +
  \sum_{j\neq k}
  \mathbf{h}_k(\boldsymbol{\ell})^{\rm H}\mathbf{w}_j s_j
  +
  z_k,
  \label{eq:fas_rx}
\end{equation}
where \(\mathbf{w}_k\in\mathbb{C}^{N_{\rm t}\times1}\) is the beamforming
vector for Rx \(k\), \(s_k\) is the corresponding data symbol, and
\(z_k\sim\mathcal{CN}(0,\sigma^2)\) denotes additive noise.
Accordingly, the SINR of Rx \(k\) under the Tx port configuration
\(\boldsymbol{\ell}\) is given by
\begin{equation}
  \gamma_k(\boldsymbol{\ell})
  =
  \frac{
  \left|\mathbf{h}_k(\boldsymbol{\ell})^{\rm H}\mathbf{w}_k\right|^2
  }{
  \sum_{j\neq k}
  \left|\mathbf{h}_k(\boldsymbol{\ell})^{\rm H}\mathbf{w}_j\right|^2
  +\sigma^2
  }.
  \label{eq:fas_sinr}
\end{equation}
The active Tx port configuration can be selected according to the network
utility of interest, e.g., sum rate, weighted sum rate, or max-min fairness.
For example, a utility-driven port configuration is written as
\begin{equation}
  \boldsymbol{\ell}^{\star}
  \in
  \arg\max_{\boldsymbol{\ell}\in\{1,\ldots,N\}^{N_{\rm t}}}
  U\!\left(\gamma_1(\boldsymbol{\ell}),\ldots,
  \gamma_K(\boldsymbol{\ell})\right),
  \label{eq:fas_port_selection}
\end{equation}
where \(U(\cdot)\) denotes the adopted downlink utility function. The key
feature of FAS is that the effective channel vector
\(\mathbf{h}_k(\boldsymbol{\ell})\) changes with the active Tx port
configuration due to port-dependent spatial fading. The system exploits
such port-domain channel variation by jointly selecting
\(\boldsymbol{\ell}\) and designing the Tx beamforming vectors
\(\{\mathbf{w}_k\}_{k=1}^{K}\). Hence, the joint optimization of Tx port
configuration and beamforming is a central design challenge in FAS-enabled
multi-Rx downlink systems.

\subsubsection{Spatial Correlation and Channel Model}

The spatial correlation among FAS ports determines how much channel
diversity can be obtained by changing the active Tx port configuration.
For the \(m\)-th Tx radiating element, let \(h_{k,m}^{(n)}\) denote the
scalar channel from its \(n\)-th port to Rx \(k\), where
\(n\in\{1,\ldots,N\}\). Given the Tx port configuration
\(\boldsymbol{\ell}=[\ell_1,\ldots,\ell_{N_{\rm t}}]^{\top}\), the
effective channel vector from the Tx-FAS array to Rx \(k\) is
\begin{equation}
  \mathbf{h}_k(\boldsymbol{\ell})
  =
  [h_{k,1}^{(\ell_1)},\ldots,
  h_{k,N_{\rm t}}^{(\ell_{N_{\rm t}})}]^{\top}
  \in\mathbb{C}^{N_{\rm t}\times1}.
  \label{eq:fas_active_channel}
\end{equation}

For uniformly spaced ports within a compact one-dimensional FAS aperture,
the correlation between ports \(m\) and \(n\) of the same radiating element
can be modeled under Clarke's isotropic scattering model as~\cite{new2024tutorial}
\begin{equation}
  \rho_{mn}
  =
  J_0\!\left(\frac{2\pi |m-n|\Delta}{\lambda}\right)
  =
  J_0\!\left(2\pi |m-n|\frac{W_s}{N-1}\right),
  \label{eq:fas_corr}
\end{equation}
where \(J_0(\cdot)\) is the zeroth-order Bessel function of the first kind,
\(\lambda\) is the carrier wavelength, \(W_s\) is the normalized FAS
aperture length in wavelengths, and
\(\Delta=W_s\lambda/(N-1)\) is the physical inter-port spacing. The
corresponding port correlation matrix is $\mathbf{R}_{p}=[\rho_{mn}]_{m,n=1}^{N}
  \in\mathbb{C}^{N\times N}$. For the \(m\)-th Tx radiating element and Rx \(k\), we define the port-domain
channel vector as
\begin{equation}
  \mathbf{c}_{k,m}
  =
  [h_{k,m}^{(1)},\ldots,h_{k,m}^{(N)}]^{\top}
  \in\mathbb{C}^{N\times1}.
  \label{eq:fas_port_channel_vector}
\end{equation}
A simple correlated Rayleigh model can then be written as $\mathbf{c}_{k,m}\sim\mathcal{CN}\!\left(\mathbf{0},\beta_{k,m}\mathbf{R}_{p}\right)$, where \(\beta_{k,m}\) denotes the large-scale fading coefficient between
the \(m\)-th Tx radiating element and Rx \(k\). Equivalently, the
cross-correlation between two-port channels of the same Tx radiating
element is
\begin{equation}
  \mathbb{E}\!\left[
  h_{k,m}^{(a)} h_{k,m}^{(b)*}
  \right]
  =
  \beta_{k,m}\rho_{ab}.
  \label{eq:fas_scalar_channel_corr}
\end{equation}

The model shows that highly correlated ports provide similar channel
coefficients and therefore limited port-selection diversity. Larger
effective aperture size \(W_s\), or larger inter-port spacing, generally
reduces the correlation and provides more distinguishable channel
realizations across candidate ports. In practice, the benefit is limited by
the physical aperture size, mutual coupling, switching overhead, and the
available channel observations.

Based on the SINR in \eqref{eq:fas_sinr}, an Rx-specific outage benchmark
under Tx port selection can be written as
\begin{equation}
  P_{{\rm out},k}
  =
  \Pr\!\left(
  \max_{\boldsymbol{\ell}\in\{1,\ldots,N\}^{N_{\rm t}}}
  \gamma_k(\boldsymbol{\ell})
  <
  \gamma_0
  \right),
  \label{eq:fas_outage}
\end{equation}
where \(\gamma_0\) is the target SINR threshold. In a multi-Rx downlink,
a network usually determines the selected Tx port configuration
utility, such as sum rate, weighted sum rate, or max-min fairness, rather
than by the outage of one Rx alone. The correlation model in
\eqref{eq:fas_corr} assumes rich isotropic scattering and uniformly spaced
ports. For sparse multipath propagation or irregular port layouts,
geometry-aware or field-response channel models may provide a more accurate
description~\cite{new2024tutorial}.

\subsubsection{Fluid Antenna Multiple Access}

A distinctive feature of FAS is its ability to enable FAMA, in which port diversity is exploited to suppress
multi-user interference without explicit coordination among Rxs. In FAMA, each Rx independently selects a port that observes weak aggregate
interference while maintaining sufficient desired-signal strength from the
serving Tx~\cite{waqar2024opportunistic}. With sufficient port density $N$ and spatial richness, the probability of finding such a port approaches 1, enabling near-interference-free reception for all Rxs simultaneously. Slow FAMA~\cite{waqar2023deep} extends this to frequency-selective fading, while opportunistic FAMA (O-FAMA) further exploits temporal variations to enhance throughput~\cite{waqar2024opportunistic}. FAMA represents a fundamentally novel multiple access paradigm with no counterpart in MA or PASS.
\subsubsection{Unique Advantages of Fluid Antenna Systems}

%FAS offers a set of physically distinct advantages that address key challenges in wireless networks across six dimensions.

\paragraph{Spatial Diversity Without Mechanical Movement}
The most fundamental advantage of FAS is its ability to achieve large
spatial diversity gains within an extremely compact form factor, without
any mechanical movement. By electronically switching among $N$ ports
spanning a space of $W_s\lambda$, FAS captures spatial channel variation
at a speed limited only by switching electronics, being orders of magnitude
faster than mechanical MA repositioning~\cite{new2024tutorial,liu2026switching}.
The best-port selection achieves a signal-to-noise ratio (SNR) distribution that stochastically
dominates single-antenna reception, with diversity order growing with respect to both
$N$ and $W_s$. This characteristic makes FAS uniquely suited to compact IoT devices and
mobile handsets where mechanical reconfigurability is impractical.

\paragraph{Interference Exploitation via FAMA}
Although the preceding model focuses on Tx-side FAS reconfiguration,
many FAMA studies consider receiver-side port selection, where each Rx is
equipped with an FAS. In receiver-side FAMA, each Rx independently selects
a port with a low aggregate interference level while maintaining sufficient
desired-signal strength from the serving Tx, enabling low-interference
reception without centralized interference coordination or successive
interference cancellation under rich scattering
~\cite{waqar2024opportunistic,waqar2023deep}. A larger effective FAS
aperture, or a set of more decorrelated ports, increases the chance of
finding a favorable receive port. Opportunistic FAMA further combines user
scheduling and port selection to improve FAMA performance in dense
multi-Rx scenarios~\cite{waqar2024opportunistic}.

\subsection{Pinching Antenna System}
% TODO: Waveguide structure, dielectric particles, pinching beamforming

\subsubsection{Architecture and Operating Principle}

Pinching antenna systems represent a fundamentally different paradigm for
spatial antenna reconfiguration. A PASS consists of one or more flexible
dielectric waveguides of arbitrary length, routed throughout a coverage
area and connected to the Tx via RF chains. Small dielectric particles,
referred to as PAs, are attached to or
detached from the waveguide surface at arbitrary positions, creating
localized electromagnetic coupling points that radiate the guided signal
into free space~\cite{xu2026joint}. The key benefit is that PAs can be deployed along waveguides closer to
Rxs than conventional fixed-position antennas, when the waveguide layout
permits. Such proximity shortens the wireless propagation distance and
helps mitigate large-scale path loss, which is a fundamental limitation of
conventional fixed-position antenna systems.
We consider a PASS-enabled downlink system with \(W\) waveguides of length \(L\) and
\(M\) PAs per waveguide serves \(K\) single-antenna Rxs. The position of the
$m$-th PA on waveguide $w$ is determined by its longitudinal coordinate
$d_{w,m} \in [0, L]$, giving the physical location
$\mathbf{q}_{w,m} = \mathbf{q}_{w,0} + d_{w,m}\hat{\mathbf{e}}_w$, where
$\mathbf{q}_{w,0}$ is the waveguide anchor at the Tx and $\hat{\mathbf{e}}_w$
is the unit direction vector of waveguide $w$. A minimum inter-PA spacing
$d_{\min}$ is enforced on each waveguide to prevent electromagnetic coupling
between adjacent PAs~\cite{xu2026joint}.

\subsubsection{Channel Model and Received Signal}

By combining the in-waveguide propagation phase and the free-space LoS propagation from the PA to the Rx, the effective PASS coefficient from the input of waveguide \(w\) to the Rx
\(k\) through the \(m\)-th PA can be formulated as \cite{xu2026joint}
\begin{equation}
  g_{w,m,k}
  =
  \underbrace{
  e^{-j\frac{2\pi}{\lambda_g}d_{w,m}}
  }_{\text{in-waveguide phase}}
  \underbrace{
  \frac{\lambda}{4\pi r_{w,m,k}}
  e^{-j\frac{2\pi}{\lambda}r_{w,m,k}}
  }_{\text{free-space LoS channel}},
  \label{eq:pass_channel}
\end{equation}
where
where \(\mathbf{q}_{w,m}\) and \(\mathbf{u}_k\) denote the positions of the
\(m\)-th PA on waveguide \(w\) and Rx \(k\), respectively, and
\(r_{w,m,k}=\|\mathbf{q}_{w,m}-\mathbf{u}_k\|_2\) is the corresponding
Euclidean distance. The variable \(d_{w,m}\) denotes the propagation
distance from the input of waveguide \(w\) to the \(m\)-th PA along the
waveguide. Moreover, \(\lambda\) is the free-space wavelength and
\(\lambda_g=\lambda/n_{\rm eff}\) is the guided wavelength, where
\(n_{\rm eff}\) denotes the effective refractive index of the dielectric
waveguide. For a simple dielectric approximation, \(n_{\rm eff}\) can be
approximated by \(\sqrt{\epsilon_r}\), with \(\epsilon_r\) being the relative
permittivity~\cite{xu2026joint,guo2025graph}. The same \(e^{-j(\cdot)}\)
phase convention is used for both the free-space and in-waveguide
propagation terms. The free-space factor \(\lambda/(4\pi r_{w,m,k})\)
captures the distance-dependent LoS attenuation, while the in-waveguide
phase term accounts for the phase accumulated before the signal reaches the
PA attachment point. The model follows a lossless-waveguide approximation.
When waveguide attenuation is non-negligible, an additional amplitude
attenuation factor depending on \(d_{w,m}\) can be included.

Collecting all \(WM\) PA-to-Rx-\(k\) channel coefficients into the
column vector \(\mathbf{g}_k\in\mathbb{C}^{WM\times1}\), the Tx transmits
the superimposed signal \(\mathbf{x}=\sum_{k=1}^{K}\mathbf{w}_k s_k\),
where \(\mathbf{w}_k\in\mathbb{C}^{WM\times1}\) is the beamforming vector
for Rx \(k\), and \(s_k\sim\mathcal{CN}(0,1)\) is the data symbol. 
The
received signal at Rx $k$ is expressed as
\begin{equation}
  y_k = \mathbf{g}_k^{\rm{H}} \mathbf{w}_k s_k
      + \sum_{j \neq k} \mathbf{g}_k^{\rm{H}} \mathbf{w}_j s_j + z_k,
  \label{eq:pass_rx}
\end{equation}
where $z_k \sim \mathcal{CN}(0, \sigma^2)$ is additive white Gaussian
noise. The SINR at user $k$ is given by
\begin{equation}
  \gamma_k = \frac{|\mathbf{g}_k^{\rm{H}} \mathbf{w}_k|^2}
    {\sum_{j \neq k}|\mathbf{g}_k^{\rm{H}} \mathbf{w}_j|^2 + \sigma^2}.
  \label{eq:pass_sinr}
\end{equation}
Compared with MA, PASS is often modeled with a more structured channel:
each PA-to-Rx link is mainly characterized by a near-field LoS propagation
path, while the signal also accumulates a guided-wave phase before reaching
the PA, as shown in~\eqref{eq:pass_channel}. In general, joint PA-position
and beamforming design in PASS remains non-convex. For example, maximizing
the sum rate \(\sum_k\log_2(1+\gamma_k)\) over the PA positions
\(\{d_{w,m}\}\) and beamforming vectors \(\{\mathbf{w}_k\}\) is difficult
because the PA positions are coupled with the channel coefficients in
\eqref{eq:pass_channel}, motivating AI-based optimization methods
~\cite{xu2026joint,guo2025graph,xu2025beamforming}. Channel estimation is
also challenging because the coefficients \(\{g_{w,m,k}\}\) depend on both
near-field PA--Rx distances and in-waveguide phases, which motivates
DL-based estimators~\cite{xiao2025channel,lv2026deep}.

\begin{table*}[!t]
\centering
\caption{Architectural and Channel-Level Comparison of MA, FAS, and PASS}
\label{tab:antenna_comparison}
\renewcommand{\arraystretch}{1.2}
\setlength{\tabcolsep}{1.5pt}
\scriptsize

\begin{tabular}{|l|l|l|l|}
\hline
\cellcolor{myblue}\tcell{2.0cm}{\textbf{Feature}} &
\cellcolor{myblue}\tcell{5.0cm}{\textbf{Movable Antenna (MA)}} &
\cellcolor{myblue}\tcell{5.0cm}{\textbf{Fluid Antenna System (FAS)}} &
\cellcolor{myblue}\tcell{5.0cm}{\textbf{Pinching Antenna System (PASS)}} \\
\hline

\tcell{2.0cm}{\textbf{Reconfigurable unit}} &
\tcell{5.0cm}{Physical antenna elements with adjustable positions} &
\tcell{5.0cm}{Radiating element or RF chain selecting among predefined ports} &
\tcell{5.0cm}{Pinching antennas attached to dielectric waveguides} \\
\hline

\tcell{2.0cm}{\textbf{Reconfiguration}} &
\tcell{5.0cm}{Continuous movement within a bounded local region} &
\tcell{5.0cm}{Discrete switching among ports within a compact aperture} &
\tcell{5.0cm}{Placement of PAs along one or multiple waveguides} \\
\hline

\tcell{2.0cm}{\textbf{Control variable}} &
\tcell{5.0cm}{Tx/Rx antenna positions, e.g., \(\mathbf{P}_{\rm t}\) and \(\mathbf{P}_{\rm r}\)} &
\tcell{5.0cm}{Port index or port configuration, e.g., \(n\) or \(\boldsymbol{\ell}\)} &
\tcell{5.0cm}{PA attachment coordinate \(d_{w,m}\in[0,L]\)} \\
\hline

\tcell{2.0cm}{\textbf{Channel basis}} &
\tcell{5.0cm}{Multipath field response with position-dependent phases} &
\tcell{5.0cm}{Port-correlated fading and port-domain spatial diversity} &
\tcell{5.0cm}{Near-field PA--Rx LoS propagation with guided-wave phase} \\
\hline

\tcell{2.0cm}{\textbf{Spatial scale}} &
\tcell{5.0cm}{Local aperture, typically several wavelengths around the device} &
\tcell{5.0cm}{Compact aperture integrated into the Tx or Rx device} &
\tcell{5.0cm}{Waveguide-level deployment over rooms, corridors, factories, or venues} \\
\hline

\tcell{2.0cm}{\textbf{Primary gain}} &
\tcell{5.0cm}{Fine-grained channel shaping and spatial multiplexing improvement} &
\tcell{5.0cm}{Port-domain diversity and interference management through port selection} &
\tcell{5.0cm}{Shorter wireless propagation distance and improved distributed coverage} \\
\hline

\tcell{2.0cm}{\textbf{CSI/observation}} &
\tcell{5.0cm}{Position-dependent CSI or channel map over candidate locations} &
\tcell{5.0cm}{Port-level CSI, port correlation, or partial port observations} &
\tcell{5.0cm}{PA--Rx near-field channel and in-waveguide phase information} \\
\hline

\tcell{2.0cm}{\textbf{Key constraint}} &
\tcell{5.0cm}{Movement latency, actuator complexity, and CSI acquisition overhead} &
\tcell{5.0cm}{Port correlation, switching cost, and port observation overhead} &
\tcell{5.0cm}{Waveguide routing, waveguide loss, and PA spacing constraints} \\
\hline

\tcell{2.0cm}{\textbf{Typical use}} &
\tcell{5.0cm}{Compact MIMO, indoor links, multipath-rich scenarios, ISAC} &
\tcell{5.0cm}{Dense access, compact terminals, receiver-side FAMA, IoT devices} &
\tcell{5.0cm}{Factories, venues, corridors, and distributed indoor Rxs} \\
\hline

\end{tabular}
\end{table*}
\subsubsection{Unique Advantages of Pinching Antenna Systems}

\paragraph{Near-User Placement and Large-Scale Path Loss Elimination}
The most distinctive advantage of PASS is its ability to place radiating
elements in the immediate vicinity of Rxs, regardless of where those
Rxs are located. Conventional systems suffer from large-scale path loss
scaling as $r^{-\alpha}$ with $\alpha \geq 2$, which dominates the link
budget in large cells. By repositioning PAs to within 1--5 meters of
Rx, PASS effectively reduces $r$ to near-zero, substantially reducing the large-scale path loss as a design constraint entirely~\cite{xu2026joint,karagiannidis2025deep}. Such proximity gain is difficult to obtain with MA, whose movement is
limited to a small local region, or with FAS, whose ports are confined
within the device aperture. PASS therefore establishes wireless links by
bringing the radiating points closer to the Rxs. The effective-rank analysis in~\cite{yang2026effective} shows that PASS can achieve higher spatial multiplexing gains than MA and FPA for a given number of radiating elements, mainly because PA placement can shorten the wireless propagation distance and mitigate the path-loss bottleneck.

\paragraph{Unlimited Coverage via Waveguide Extension}
Unlike MA, which is confined to a movable region of a few wavelengths, and
FAS, which is limited to a compact device form factor, PASS waveguides can
extend over arbitrary lengths, potentially tens or hundreds of meters, to cover an entire room, factory floor, or campus
area~\cite{xu2026joint,wu2026straggler}. This scalability makes PASS
uniquely suited for wide-area deployments with highly distributed Rxs,
such as industrial IoT, smart factories, and indoor
hotspots~\cite{kang2025campass}. New Rxs entering the coverage area
can be served by dynamically pinching a PA at the nearest waveguide
segment, without any infrastructure changes.

\paragraph{Flexible ISAC via Reconfigurable Aperture}
By repositioning PAs along the waveguide, PASS dynamically controls the
effective aperture and geometry of the transmit array. This reconfigurability
provides an additional DoF for ISAC: PA positions can be jointly optimized
with transmit power and beamforming to maximize communication simultaneously
SINR and minimize the CRB for target localization~\cite{qin2025joint}. Segmented waveguide architectures further enable
independent control of different waveguide sections, allowing simultaneous
multi-function ISAC operation~\cite{gao2026integrated,gao2026llm}.

\paragraph{Physical Layer Security and Covert Communications}
PASS provides a natural and powerful mechanism for physical layer security:
by placing PAs close to the legitimate user while keeping them far from
potential eavesdroppers, the system creates an inherent signal asymmetry
that is difficult to overcome without knowledge of the PA
positions~\cite{jiang2026pinching}. In sensing-aided covert communications,
PASS dynamically tracks the warden's position via sensing and repositions
PAs to suppress the warden's received power below the noise detection
threshold, while maintaining covert throughput to the legitimate
receiver~\cite{jiang2026pinching}.% This sensing-aided dynamic repositioning capability has no counterpart in fixed-array covert communication systems.

\iffalse
\paragraph{Latency Reduction and Straggler Mitigation in Distributed Learning}
PASS dramatically reduces communication latency in wireless federated
learning (FL) by dynamically establishing strong near-field LoS links to
the slowest-uploading devices (stragglers)~\cite{wu2026straggler}. By
repositioning PAs toward straggler devices, the hybrid conventional and
pinching antenna network (HCPAN) ensures all devices can upload model
updates within the global aggregation round deadline, eliminating the
straggler bottleneck that limits convergence of distributed
learning~\cite{wu2026straggler}. In NOMA-mobile edge computing (MEC) systems, PASS can reduce task
offloading latency by improving the effective offloading channel from the
Tx to each Rx, thereby enabling faster computation offloading and lower
end-to-end service latency~\cite{ma2026joint}.

Table~\ref{tab:antenna_comparison} summarizes the main architectural and
channel-level differences among MA, FAS, and PASS. Although all three
technologies introduce spatial reconfigurability into antenna systems, they
operate at different physical scales and rely on different reconfiguration
mechanisms. MA continuously adjusts antenna positions within local movable
regions. FAS selects discrete ports within a compact aperture at the Tx or
Rx. PASS places radiating points along dielectric waveguides to shorten the
wireless propagation distance to distributed Rxs. As a result, the three
architectures differ in channel modeling, CSI acquisition, hardware
constraints, and suitable deployment scenarios.
\fi

\begin{table*}[!t]
\centering
\caption{Comparison of AI Design Implications Across MA, FAS, and PASS}
\label{tab:ai_comparison}
\renewcommand{\arraystretch}{1.2}
\setlength{\tabcolsep}{1.5pt}
\scriptsize

\begin{tabular}{|l|l|l|l|}
\hline
\cellcolor{myblue}\tcell{2.0cm}{\textbf{AI Aspect}} &
\cellcolor{myblue}\tcell{5.0cm}{\textbf{Movable Antenna (MA)}} &
\cellcolor{myblue}\tcell{5.0cm}{\textbf{Fluid Antenna System (FAS)}} &
\cellcolor{myblue}\tcell{5.0cm}{\textbf{Pinching Antenna System (PASS)}} \\
\hline

\tcell{2.0cm}{\textbf{State input}} &
\tcell{5.0cm}{Position-dependent CSI, Tx/Rx position matrices, spatial statistics} &
\tcell{5.0cm}{Port CSI, port correlation, switching history} &
\tcell{5.0cm}{PA coordinates, Rx geometry, guided-wave phase} \\
\hline

\tcell{2.0cm}{\textbf{Action}} &
\tcell{5.0cm}{Continuous position vectors and beamforming} &
\tcell{5.0cm}{Discrete port indices and continuous beamforming} &
\tcell{5.0cm}{Continuous PA positions and beamforming} \\
\hline

\tcell{2.0cm}{\textbf{Optimization type}} &
\tcell{5.0cm}{Continuous nonconvex co-design} &
\tcell{5.0cm}{Mixed discrete--continuous optimization} &
\tcell{5.0cm}{Geometry-constrained continuous co-design} \\
\hline

\tcell{2.0cm}{\textbf{Learning target}} &
\tcell{5.0cm}{Rate, sensing, security, spacing feasibility} &
\tcell{5.0cm}{Throughput, outage, switching cost, FAMA performance} &
\tcell{5.0cm}{Rate, sensing, proximity gain, PA spacing feasibility} \\
\hline

\tcell{2.0cm}{\textbf{Typical AI tools}} &
\tcell{5.0cm}{DDPG, TD3, SAC, deep unfolding, meta-RL} &
\tcell{5.0cm}{DQN, MARL, GNN, Transformer, LLM} &
\tcell{5.0cm}{SAC, DDPG, GNN, KKT-guided learning} \\
\hline

\tcell{2.0cm}{\textbf{Network structure}} &
\tcell{5.0cm}{DNN, unfolded WMMSE, position prediction network} &
\tcell{5.0cm}{Port-Rx GNN, port-sequence Transformer, LLM-guided framework} &
\tcell{5.0cm}{Waveguide graph network, projection layer, KKT-guided DNN} \\
\hline

\tcell{2.0cm}{\textbf{Main challenge}} &
\tcell{5.0cm}{Continuous exploration, CSI sampling, position-beamforming coupling} &
\tcell{5.0cm}{Combinatorial port search, port correlation, switching overhead} &
\tcell{5.0cm}{Near-field geometry, waveguide phase, multi-waveguide coordination} \\
\hline

\tcell{2.0cm}{\textbf{Scalability strategy}} &
\tcell{5.0cm}{Two-timescale control, implicit CSI, meta-RL} &
\tcell{5.0cm}{Candidate pruning, mean-field RL, sparse port observations} &
\tcell{5.0cm}{Hierarchical control, graph generalization, structural constraints} \\
\hline

\tcell{2.0cm}{\textbf{Representative work}} &
\tcell{5.0cm}{\cite{liang2026two,weng2024learning,xiu2026meta}} &
\tcell{5.0cm}{\cite{guo2025llm,he2025graph,waqar2024opportunistic}} &
\tcell{5.0cm}{\cite{xu2026joint,guo2025graph,xu2025beamforming}} \\
\hline
\end{tabular}
\vspace{-2em}
\end{table*}

\subsection{Challenges of Conventional Optimization and the Necessity of AI}
%------------------------------------------------------------------
%The joint optimization problems associated with MA, FAS, and PASS impose several challenges that make conventional optimization methods as follows. Meanwhile, AI models have demonstrated remarkable potential to address these problems, in that they are able to learn efficient mappings from system states to high-quality control decisions. 

\paragraph{Non-convexity and Coupled Variables} In MA, the channel is a
nonlinear trigonometric function of antenna positions through
\eqref{eq:ma_channel}, making the joint position-beamforming problem
non-convex with many local optima~\cite{liang2026two}. In FAS, the joint
port selection and beamforming problem is a mixed-integer non-convex program
where the coupling between discrete port indices and continuous beamforming
weights makes even approximate solutions NP-hard. In PASS, the waveguide
phase in~\eqref{eq:pass_channel} introduces complex exponential coupling
between PA positions and beamforming, thwarting successive convex approximation (SCA) and alternating optimization (AO) methods without strong initialization. AI techniques, including supervised, unsupervised, and self-supervised learning, DRL, and deep unfolding, can effectively address this challenge. For example, unsupervised or self-supervised learning can directly map the channel state to antenna configuration and beamforming variables, with physical constraints incorporated through penalty terms or projection layers~\cite{kang2025campass}, thereby avoiding repeated online optimization. Meanwhile, DRL algorithms search for optimal policies through exploration, allowing the agent to examine a broader solution space and reduce its dependence on deterministic local search trajectories~\cite{weng2024learning}.

% \paragraph{High dimensionality and combinatorial complexity} For FAS with
% $N=64$ ports per Rx and $K=8$ Rxs, the joint port selection space
% reaches $64^8 \approx 2.8 \times 10^{14}$ combinations~\cite{new2024tutorial}.
% For PASS with $W=4$ waveguides and $M=8$ PAs each, the 32-dimensional
% continuous position space coupled with beamforming creates an intractable
% optimization landscape. For 6DMA systems, antenna position, orientation,
% and beamforming are optimized jointly over a manifold of dimension
% $6N + 2N_{\rm t}K$~\cite{shao2025hybrid}. 

\paragraph{High Dimensionality and Combinatorial Complexity} In MA, FAS, and PASS systems, optimization problems are typically high-dimensional, possibly involving extremely large search spaces. For example, a FAS system with $N = 64$ ports per Rx and $K = 8$ Rxs, the joint port selection space contains $64^8 \approx 2.8 \times 10^{14}$ combinations \cite{new2024tutorial}. Similarly, 6DMA requires the joint optimization of antenna position, orientation, and beamforming variables over a space of dimension $6N + 2N_{\rm t}K$ \cite{shao2025hybrid}. Thus, traditional algorithms such as exhaustive and branch-and-bound searches may become impractical within a channel coherence interval, while grid-based and iterative optimization methods become increasingly computationally demanding. AI techniques can address this challenge by replacing repeated online search with learned inference. For example, for the combinatorial port selection problem in FAS, LLM-based hyper-heuristics can generate promising candidate configurations without exploring the entire search space and adapt to new problem instances through few-shot learning~\cite{guo2025llm, wang2025llm}.

\paragraph{Dynamic Environments and Real-time Constraints} Conventional
iterative solvers, such as weighted minimum mean square error (WMMSE) and AO, must re-solve the full problem at every coherence interval, which is intractable in mobile settings. For MA, the CSI at a candidate position must be acquired before repositioning, creating a fundamental chicken-and-egg problem~\cite{11500467}. For FAS in fast-fading environments, port selection must be executed within the
coherence time, which is far too fast for iterative solvers to converge. AI techniques alleviate this burden through offline-trained inference, where most computation is performed offline during training and online decisions are generated through a forward pass~\cite{kang2025campass}. For example, DRL learns a policy that directly maps the current observation to an action, thereby avoiding repeated online optimization~\cite{weng2024learning}. When the operating environment differs from the training distribution, meta-learning can provide an effective initialization that enables rapid adaptation with only a few update steps~\cite{liang2025two}.

% \paragraph{Imperfect CSI and robustness} Robust formulations accounting for
% CSI uncertainty lead to min-max or stochastic problems significantly more
% demanding than their nominal counterparts~\cite{xiu2026robust}.
% %These challenges collectively motivate AI as the natural solution: DRL learns near-optimal policies from environment interaction without re-solving the optimization problem at each step; deep learning amortizes optimization cost across instances; and LLMs provide combinatorial reasoning for the port selection challenge in FAS~\cite{guo2025llm,wang2025llm}.
% These challenges do not invalidate classical optimization methods, which
% remain essential for benchmarking, initialization, and label generation.
% However, they motivate AI as a complementary and increasingly important
% framework for learning fast, adaptive, and low-complexity policies~\cite{guo2025llm,wang2025llm}.

\paragraph{CSI Acquisition and Uncertainty} A challenge in spatially reconfigurable antenna systems is that the CSI depends on the selected antenna configuration, while selecting the appropriate configuration itself requires knowledge of the CSI. For example, in MA systems, the CSI at candidate positions must be acquired before determining the antenna locations \cite{11500467}, whereas in FASs, estimating the CSI across all available ports may incur significant pilot overhead. Furthermore, the acquired CSI is naturally affected by estimation errors, and robust formulations that account for such uncertainty often lead to computationally demanding stochastic or worst-case optimization problems \cite{xiu2025movable}. AI techniques can reduce both the CSI acquisition overhead and the impact of imperfect CSI, where data-driven extrapolation schemes reconstruct the full position or port-domain CSI from a limited set of observations \cite{zhang2024learning, gao2025ssnet}, while implicit-CSI methods directly map received-signal features to antenna configurations without explicitly estimating the complete channel~\cite{lu2025learning}.

%------------------------------------------------------------------
\subsection{Fundamental Differences Among AI Models for MA, FAS, and PASS}
%------------------------------------------------------------------
 The specific physical layer characteristics of each antenna system, i.e., MA,
FAS, or PASS,
fundamentally shape how AI models are used as summarized in Table~\ref{tab:ai_comparison}.
%and~\ref{tab:ai_algo_summary}.

\subsubsection{State Space and Physical Layer Inputs}
An AI model typically take physical-layer quantities as input for making its decisions, and they are different accross the antenna architectures.
For MA, the state encodes position-dependent channel information:
the field-response matrix at candidate positions, the current TPV, and
local spatial statistics. Compact implicit CSI representations are preferred
over full channel maps to reduce state dimensionality~\cite{lu2025learning}.
For FAS, the state is naturally port-indexed: channel estimates
at each port, the spatial correlation matrix, and port switching history
(since switching costs must be penalized)~\cite{guo2025llm}. For PASS, the
state encodes waveguide geometry and Rx proximity: PA positions,
Rx location estimates, and in-waveguide phase~\cite{xu2026joint}.
 
\subsubsection{Action Space Structure}
The action space is defined by the optimization variables, which is thus different accross antenna systems. In MA, the action is fully continuous over the TPV, jointly with
continuous beamforming, favoring deep deterministic policy gradient (DDPG), soft actor-critic (SAC), and twin delayed DDPG (TD3)~\cite{weng2024learning}.
In FAS, the mixed discrete-continuous structure motivates a deep Q-network (DQN) for port
selection combined with neural beamforming, or LLM-based hyper-heuristics
that handle the combinatorial search~\cite{guo2025llm,wang2025llm}. In PASS,
the action is continuous but geometrically constrained along the waveguide,
requiring projection layers to enforce minimum PA spacing~\cite{xu2026joint}.
 
\subsubsection{Reward Function Design}
For MA, rewards combine sum-rate with penalties on spacing violations and
region boundaries~\cite{liang2026two}. For FAS, the reward incorporates
port switching costs alongside throughput~\cite{liu2026switching}; FAMA
rewards uniquely incentivize selecting ports coinciding with interferer deep
fades~\cite{waqar2024opportunistic}. For PASS, rewards capture proximity
gain and penalize waveguide length violations~\cite{ma2026joint}.
 
\subsubsection{Neural Network Architecture}
For MA, deep neural networks (DNNs) and deep unfolding networks embedding WMMSE achieve
strong performance with interpretability~\cite{liang2025two,feng2026deep}.
For FAS, GNNs exploit the port-Rx graph structure, Transformers handle
long-range port dependencies, and LLMs address the combinatorial port
space~\cite{he2025graph,guo2025llm}. For PASS, GNNs naturally model the
waveguide topology, and Karush--Kuhn--Tucker (KKT)-guided architectures embed optimality conditions as structural constraints~\cite{guo2025graph,xu2025beamforming}.

\iffalse
\subsubsection{Training Methodology and Sample Efficiency}
MA benefits from two-timescale approaches separating slow
repositioning from fast beamforming~\cite{liang2026two,liang2025two}. FAS
addresses combinatorial exploration via mean-field RL and LLM few-shot
generalization~\cite{gu2025partially,guo2025llm}. PASS benefits from the
smoothly varying near-field line-of-sight (LoS) channel enabling faster gradient convergence,
while KKT-guided learning reduces sample complexity~\cite{kang2025campass,xu2025beamforming}.

\fi

%=====================================================================
%=====================================================================
\section{AI for Movable Antenna Systems}
\label{sec:ai_for_MAs}
%=====================================================================

This section reviews AI-assisted antenna-position optimization in MA-enabled wireless systems. By enabling antenna elements to adapt their positions to channel, service, and environmental conditions, MAs provide additional spatial degrees of freedom. This flexibility allows antenna positions to be jointly optimized with beamforming, channel estimation, sensing, trajectory/phase-shift design, security, and resource allocation. However, the resulting problems are often non-convex, high-dimensional, and time-varying, making conventional iterative methods difficult to apply in real time. Therefore, AI techniques, including DL, DRL, MARL, meta-learning, Transformers, and deep unfolding, are increasingly used to learn efficient mappings from wireless environments to antenna configurations and resource-control decisions. The relevant studies are summarized in Tables \ref{tab:SecIII-1} and \ref{tab:SecIII-2}.

\begin{table*}[t]
\centering
\caption{Summary of AI for Movable Antenna Systems: Part I}
\label{tab:SecIII-1} 
\renewcommand{\arraystretch}{1.3}
\footnotesize
\begin{tabular}{|l|l|l|l|l|l|}
\hline
\rowcolor{myblue}
\textbf{Tax.} & \textbf{Ref.} & \textbf{Scenario} & \textbf{Objective} & \textbf{AI Technique} & \textbf{AI Role} \\
\hline

\multirow{12}{*}{\rotatebox{90}{\makecell[c]{Joint Antenna Position\\ \& BF Optimization}}}
& \cite{kang2024deep} & MA multicast & Min-gain maximization & Unsupervised DL & APV prediction \\ \cline{2-6}
& \cite{xie2024deep} & 2D MISO & SE maximization & CNN-based learning & Flexible precoding \\ \cline{2-6}
& \cite{kim2025deep} & MU-MISO & Sum-rate maximization & DNN-based learning & Position/BF prediction \\ \cline{2-6}
& \cite{weng2024learning} & Multi-receiver MA & Sum-rate maximization & Multi-agent DRL & Adaptive control \\ \cline{2-6}
& \cite{wang2026learning} & Discrete MA downlink & Sum-rate maximization & GNN-attention learning & Mixed-variable learning \\ \cline{2-6}
& \cite{liang2025two} & Two-timescale MA & Utility maximization & Deep unfolding & Low-complexity update \\ \cline{2-6}
& \cite{liang2026two} & Two-timescale MA & Sum-rate maximization & Deep unfolding + meta-learning & Online adaptation \\ \cline{2-6}
& \cite{dai2025movable} & HST CF MIMO & Uplink SE maximization & PPO-based DRL & Mobility-aware positioning \\ \cline{2-6}
& \cite{li2025deep} & CF MA & Sum-rate maximization & Multi-agent DRL + TMMSE & Distributed AP control \\ \cline{2-6}
& \cite{huang2025deep} & LEO satellite & Throughput maximization & DDQN-based DRL & Pixel control \\ \cline{2-6}
& \cite{yang2026effective} & MA/PA systems & Effective-rank maximization & Graph RL / Multi-agent graph RL & Spatial DoF learning \\ \cline{2-6}
& \cite{kang2025nmap} & Near-field XL-MIMO & Min-gain maximization & CNN-FNN hybrid DL & Near-field BF \\
\hline

\multirow{6}{*}{\rotatebox{90}{\makecell[c]{Joint Antenna \\ Position \& CE}}}
& \cite{11500467} & TDD MA & CSI acquisition & Deep-sets RL + CS & Active sensing \\ \cline{2-6}
& \cite{jang2025new} & MA CE & CE accuracy improvement & DNN-aided CE & Position/estimator design \\ \cline{2-6}
& \cite{lu2025learning} & MISO MA & Position optimization & Supervised MLP learning & Implicit CSI mapping \\ \cline{2-6}
& \cite{feng2026deep} & Wideband MA & CE + BF optimization & Transformer-aided CE/BF & CE denoising \\ \cline{2-6}
& \cite{shao2025hybrid} & Hybrid-field 6DMA & Sum-rate maximization & A2C-based DRL & 6D control \\ \cline{2-6}
& \cite{wang2026drl} & MA-OTFS & Channel gain maximization & DQN-based DRL + SBLVI & Imperfect-CSI positioning \\
\hline

\multirow{10}{*}{\rotatebox{90}{\makecell[c]{Joint Antenna \\ \&  Position Sensing}}}
& \cite{xiu2026robust} & CF-ISAC & Worst-case CRLB minimization & Meta-RL + TD3 & Robust sensing \\\cline{2-6}
& \cite{xiu2025movable} & Cooperative ISAC & Power minimization & Constrained DRL with DDPG & Constrained control \\ \cline{2-6}
& \cite{xiu2026meta} & FD CF-DFRC & WCSR maximization & Meta-RL & CFO adaptation \\ \cline{2-6}
& \cite{wang2025unsupervised} & Near-field ISAC & Rate maximization & Unsupervised DL & Constraint-aware design \\ \cline{2-6}
& \cite{zhang2026crosstalk} & Crosstalk ISAC & CRB minimization & TD3-based DRL & Crosstalk-resilient BF \\ \cline{2-6}
& \cite{amhaz2026meta} & FD ISAC & Echo SINR maximization & Gradient-based meta-learning & Fast resource update \\ \cline{2-6}
& \cite{xie2025learning} & ISCPT & Multi-objective optimization & Heterogeneous multi-agent DDPG & Flexible BF \\ \cline{2-6}
& \cite{le2025beamforming} & Secure ISAC & Secrecy-rate maximization & PPO-based DRL & Secure movement \\ \cline{2-6}
& \cite{bian2026movable} & Active RIS-ISAC & Max-min secrecy rate & AO-based D3QN-TD3 & Hybrid-action learning \\ \cline{2-6}
& \cite{gao2025distributed} & ELAA ISAC & Long-term sum-rate maximization & Dimension-reduced multi-agent PPO & Distributed selection \\
\hline

\end{tabular}
\vspace{-2em}
\end{table*}

\begin{table*}[t]
\centering
\footnotesize
\caption{Summary of AI for Movable Antenna Systems: Part II}
\label{tab:SecIII-2}  
\renewcommand{\arraystretch}{1.2} 
\begin{tabular}{|l|l|l|l|l|l|}
\hline
\rowcolor{myblue}
\textbf{Tax.} & \textbf{Ref.} & \textbf{Scenario} & \textbf{Objective} & \textbf{AI Technique} & \textbf{AI Role} \\
\hline

\multirow{4}{*}{\rotatebox{0}{\makecell[l]{Joint Antenna Position \\ \& Trajectory}}}
& \cite{ahmadzadeh2024movable} & UAV OTA-FL & MSE minimization & TD3 & AP/BF control \\\cline{2-6}
& \cite{zhou2025joint} & MA-aided A2G & Sum-rate maximization & Dual-scale DRL & Trajectory/MA control \\ \cline{2-6}
& \cite{zhou2026joint} & UAV downlink & Sum-rate maximization & TD3 + DRL & Hierarchical control \\ \cline{2-6}
& \cite{bai2025movable} & Backscatter IoT & Collection-time minimization & SAC & Trajectory/orientation control \\
\hline

\multirow{7}{*}{\makecell[l]{Joint Antenna Position \\ \& Phase Shift}}
& \cite{zhuang2025multi} & RIS-MA MISO & Sum-SE maximization & MARL & Active/passive control \\ \cline{2-6}
& \cite{yu2025learning} & STARS-MA & WSR maximization & TD3 & Full-space control \\ \cline{2-6}
& \cite{amiri2025movable} & STAR-RIS SWIPT & Rate/EH maximization & Meta-TD3 & Fast SWIPT adaptation \\ \cline{2-6}
& \cite{xie2024movable} & Covert RIS-MA & Covert-rate maximization & DRL + AO & Secure trajectory \\ \cline{2-6}
& \cite{geng2025aerial} & ME-ARIS & Sum-rate maximization & DDQN & Anti-jamming control \\ \cline{2-6}
& \cite{tang2025weighted} & Movable RIS & WSR maximization & Meta-learning & Deployment adaptation \\ \cline{2-6}
& \cite{zhang2026movable} & Vehicular MA & WCSR maximization & AO + SCA & Robust benchmark \\
\hline

\multirow{5}{*}{\makecell[l]{Joint Antenna Position\\ \& Resource Allocation}}
& \cite{liu2026movable} & URLLC & WSR maximization & TD3 & Power/position control \\ \cline{2-6}
& \cite{fang2026toward} & Uplink RSMA & Sum-rate maximization & Multi-agent TD3 & Decoding/resource control \\ \cline{2-6}
& \cite{zhang2025reinforcement} & Relay MA & Rate maximization & DDPG & Long-term relay control \\ \cline{2-6}
& \cite{amhaz2026enhancing} & CoMP-RSMA & Sum-rate maximization & Meta-learning & Large-scale optimization \\ \cline{2-6}
& \cite{amhaz2025optimizing} & C-NOMA & Sum-rate maximization & DDPG & Cooperative control \\
\hline

\multirow{5}{*}{\makecell[l]{Joint Antenna Position\\ \& Security}}
& \cite{wang2026moving} & Secure MA & Secrecy-rate improvement & Transformer & Position prediction \\ \cline{2-6}
%& \cite{li2025can} & UAV/MA PLS & ASR maximization & Optimization & Mobility benchmark \\ \cline{2-6}
& \cite{tang2025deep} & MA anti-jamming & SINR maximization & MLP & Position inference \\ \cline{2-6}
& \cite{guo2025movable} & Robust anti-jamming & Average sum-rate maximization & Model-driven DL & Convergence acceleration \\ \cline{2-6}
& \cite{zhang2025deep} & MA anti-jamming & Interference suppression & Deep unfolding & Interpretable optimization \\
\hline

\end{tabular}
\vspace{-2em}
\end{table*}

%-----------------------------------------------------------------
\subsection{Joint Antenna Position and Beamforming Optimization}
%-----------------------------------------------------------------

%\cite{huang2025deep,li2025deep,weng2024learning,kang2024deep,
%kim2025deep,xie2024deep,dai2025movable,liang2026two,liang2025two,
%wang2026learning,yang2026effective,kang2025nmap}

\begin{figure}[t]
%\vspace{-1em}
\centering
\includegraphics[width=\linewidth]{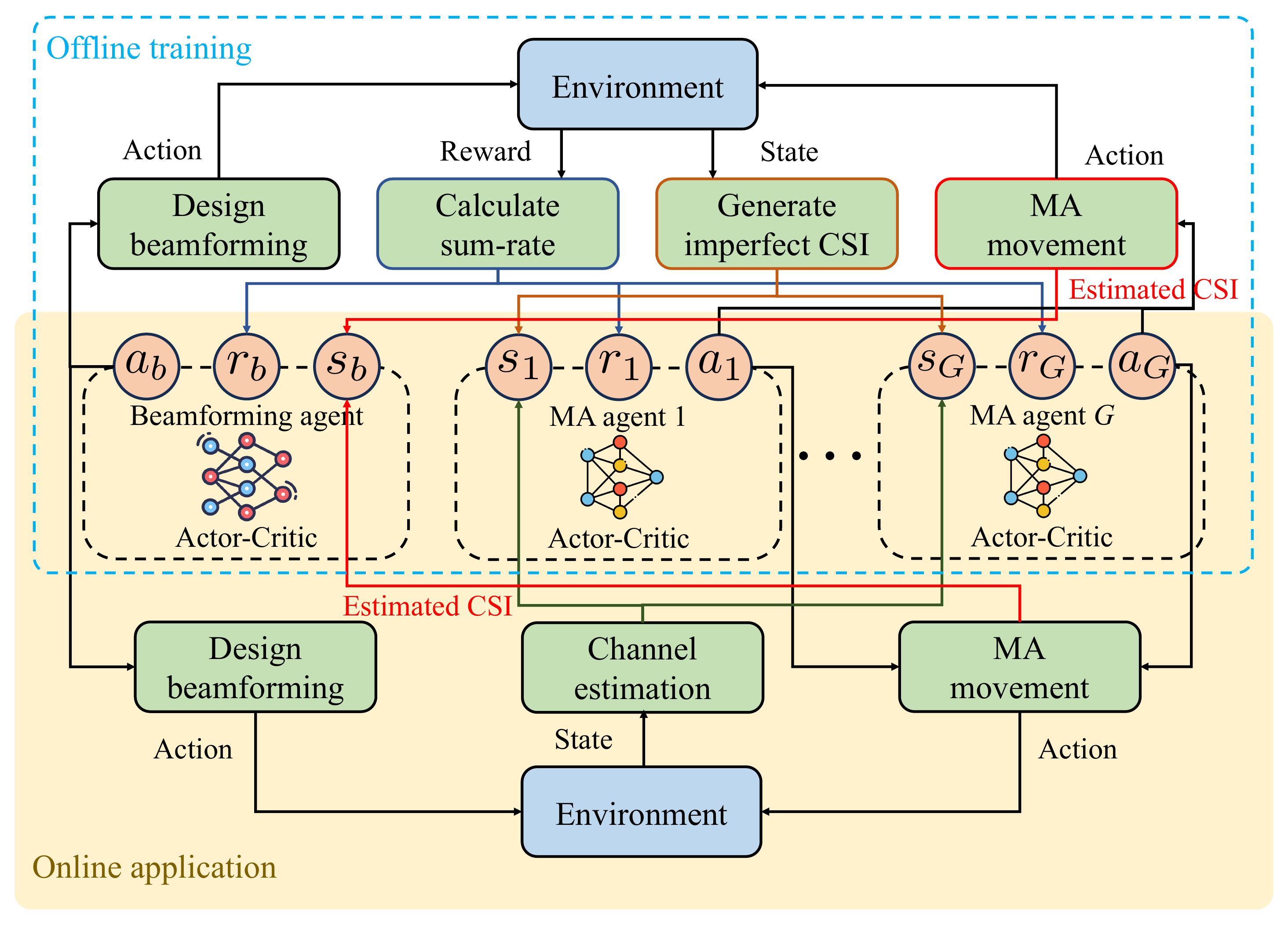}
%\vspace{-1em}
\caption{Illustration of multi-agent DRL for joint beamforming and movable antenna position optimization \cite{weng2024learning}.}
\label{Section_3_Figure4}
\vspace{-2em}
\end{figure}

Early learning-based studies mainly focus on basic multicast, multiple-input single-output (MISO), and multi-user downlink systems. In \cite{kang2024deep}, a DL-enabled multicast beamforming framework is developed for MA arrays, where the antenna position vector (APV) and antenna weight vector (AWV) are jointly optimized to maximize the minimum beamforming gain among users. AI is used to replace high-complexity iterative optimization with an unsupervised neural network that directly learns feasible antenna and beamforming configurations from channel features. The simulation results show that the proposed DL-based MA multicast beamforming scheme consistently achieves a minimum beamforming gain higher than the baselines; for example, at $N=5$, it improves the minimum beamforming gain from about $2.8$ under the strongest baseline to about $3.5$. To move toward more flexible precoding, the authors in \cite{xie2024deep} investigate a two-dimensional MISO MA system and introduce a convolutional neural network (CNN)-based framework for flexible precoding. CNN is used to extract spatial-channel features and to generate both antenna positions and beamforming weights, while the loss function incorporates the spectral-efficiency objective and antenna-spacing constraints. To further address interference-aware transmission, an MA-enabled multi-user (MU)-MISO downlink system is studied in \cite{kim2025deep}, in which two DNNs are designed to determine antenna positions and beamforming-related features, respectively. AI, therefore, serves as a low-complexity function approximator for sum-rate-oriented joint design. Simulation results show that the MA-DNN scheme achieves the highest average sum-rate, and the gain increases at high SNR; for example, at 30 dB, it achieves about 45 bps/Hz, compared with about 42 bps/Hz for FPA-DNN and about 37 bps/Hz for FPA-MRT. The authors further consider the robustness to imperfect CSI in \cite{weng2024learning}, where they jointly optimize transmit beamforming and double-sided antenna movement at both the transmitter and receiver. A heterogeneous multi-agent deep deterministic policy gradient (MADDPG) framework is introduced, as shown in Fig. \ref{Section_3_Figure4}, in which different agents, including the beamforming agent and MA transceiver agents, are responsible for beamforming and antenna movement. In this case, AI plays a decision-making role rather than only a predictive role.

Another group of studies focuses on constrained antenna positioning and practical update timescales. Joint discrete antenna positioning and continuous beamforming are formulated in \cite{wang2026learning} for a multi-user downlink system. The proposed framework combines a GNN- and attention-based encoder-decoder for discrete position selection with a GNN-based beamforming network for continuous beamforming generation. The main role of AI is to handle mixed discrete-continuous variables while satisfying constraints on coupled-antenna spacing via a masking mechanism. The simulation results show that the proposed DL framework achieves the highest sum rate and much faster inference; for example, its computation time is only $7.89$ ms, compared to $644.37$ ms for randomly selecting feasible MA positions with WMMSE precoding, and $3682.82$ ms for MA positioning with projection to discrete sampling points. In parallel, practical antenna movement also introduces a mismatch between slowly updated antenna positions and rapidly updated digital beamformers. To address this issue, a two-timescale framework is developed in \cite{liang2025two}, where the antenna position vector is optimized using long-term channel statistics, while the beamforming matrix is updated based on short-term CSI. A deep-unfolding neural network is constructed by unfolding the stochastic successive convex approximation algorithm into trainable layers. In this work, AI embeds the structure of the optimization algorithm in a readable architecture, improving efficiency while retaining interpretability. This idea is further extended in \cite{liang2026two}, where meta-learning and online adaptation are introduced to enable rapid initialization and continuous adjustment to channel statistics. Simulation results show that the proposed deep-unfolding neural network (DUNN) achieves about $122$ bits/s/Hz at $40$ dB, outperforming stochastic successive convex approximation (SSCA) at about $119$ bits/s/Hz and uniform planar array (UPA) at about $112$ bits/s/Hz; with meta-learning, the required training batches are further reduced from about $50$ to about $20$ as the number of meta-training tasks increases from $5$ to $20$.

Distributed and highly dynamic networks introduce additional challenges for joint antenna position and beamforming optimization. In \cite{dai2025movable}, MAs are incorporated into CF massive MIMO for high-speed train (HST) communications. A proximal policy optimization (PPO)-based DRL algorithm is developed to optimize antenna positions and maximize uplink spectral efficiency. AI is used to adjust multiple antenna positions in a high-mobility scenario, where the conventional AO paradigm may fail to capture rapid channel variations and Doppler effects. To jointly consider distributed antenna positioning and precoding, the authors in \cite{li2025deep} investigate MA-assisted CF networks and propose a multi-agent TD3 (MATD3)-truncated minimum mean square error (TMMSE) framework. MATD3 is used to optimize three-dimensional antenna positions at distributed access points, while TMMSE generates digital precoders with limited CSI sharing. AI therefore serves as a scalable multi-agent controller for high-dimensional antenna positioning, while model-based TMMSE reduces fronthaul overhead. Unlike terrestrial CF systems, a low-Earth-orbit (LEO) satellite system with pixel-based MAs is considered in \cite{huang2025deep}, in which antenna activation patterns and beamforming weights are jointly controlled. A double deep Q-network (DDQN) algorithm is adopted to learn adaptive decisions for long-term throughput maximization. The AI component is particularly important because the pixel-based structure creates a discrete action space, and DRL enables online antenna configuration without repeatedly solving mixed-integer optimization problems. The simulation results show that the proposed DDQN achieves an end reward of 11.3\% higher than the deep Q-network (DQN), improves throughput by 150.7\% when the number of decisions increases from $T=50$ to $T=300$, and reaches 4.23 bits / s / Hz at active elements $K=8$, compared to 2.03 bits / s / Hz at $K=2$.

Recent works also extend the discussion from rate-oriented optimization to spatial degree-of-freedom analysis and near-field beamforming. In \cite{yang2026effective}, effective rank is introduced as a structure-oriented metric to compare MA and pinching antenna systems. Instead of directly optimizing the rate or SINR, the proposed graph attention implicit quantile network (GAIQN) and multiagent graph attention Q-network (MAGAQN) are used to optimize antenna positions over multiple time slots and enhance the channel effective rank. AI is used to handle temporal decision-making, graph-structured antenna interactions, and collision-free position selection. This provides a more fundamental perspective on how flexible antenna structures exploit spatial degrees of freedom. Moving toward large-aperture systems, near-field extremely large-scale multiple-input multiple-output (XL-MIMO) communications are investigated in \cite{kang2025nmap}, where a near-field MA position network (NMAP-Net) is proposed for joint multibeamforming design and antenna position optimization. The DL architecture learns to maximize the minimum beamforming gain for desired users while suppressing interference leakage toward undesired users. Simulation results show that the proposed NMAP-Net achieves a near-upper-bound minimum beamforming gain and clearly outperforms all baselines; for example, at $N=256$ antennas, it reaches about $20$ dB, compared with about $19$ dB for Baseline I, $18.5$ dB for Baseline II, and only about $15$ dB for the fixed-position DL baseline.

%-----------------------------------------------------------------
\subsection{Joint Antenna Position and Channel Estimation}
%-----------------------------------------------------------------

%\cite{11500467,shao2025hybrid,wang2026drl,feng2026deep,jang2025new,lu2025learning}

In \cite{11500467}, a TDD MA system is considered on time-varying channels, where CS reconstructs the full CSI from limited RF-chain observations and a policy-gradient actor-critic RL algorithm selects the next sensing antenna positions. The RL state consists of previous sensing locations and complex-valued observations; the action is the sensing vector for CSI acquisition, and the reward is the achieved beamforming gain. Simulations show that RL-aided sensing consistently outperforms random sensing; when the aperture size increases to $A=12$, RL achieves about 0.85 efficiency, compared with about 0.76 for random sensing.

Moving from active sensing to joint estimator design, \cite{jang2025new} develops a learning-aided framework that jointly optimizes MA positions, AoAs, and DoAs. The received pilot model is decomposed into a neural-network-like structure, where the first layer emulates pilot construction, and its trainable weight matrix is extracted as the antenna position matrix. During training, the network maps channel angles and gains to estimated AoDs/AoAs; during inference, the optimized APVs are fixed, and the remaining layers estimate angles from received pilots. Numerical results show that larger training datasets reduce NMSE.

Another line of work reduces the overhead of channel estimation by learning MA position-control decisions from partial or implicit CSI. In \cite{lu2025learning}, a DNN-based method is proposed for MISO systems, where only the measured channel power gains at $K$ sampling points are used to predict the optimized MA-position indices. The adopted up-down MLP is trained offline with graph-based optimal positions as labels and is then used online to determine MA positions without full channel-map estimation. Numerical results show that the DNN-based scheme achieves about 95\% of the optimal received SNR as the number of MAs increases and consistently outperforms FPA systems.

\begin{figure}[t]
%\vspace{-1em}
\centering
\includegraphics[width=0.8\linewidth]{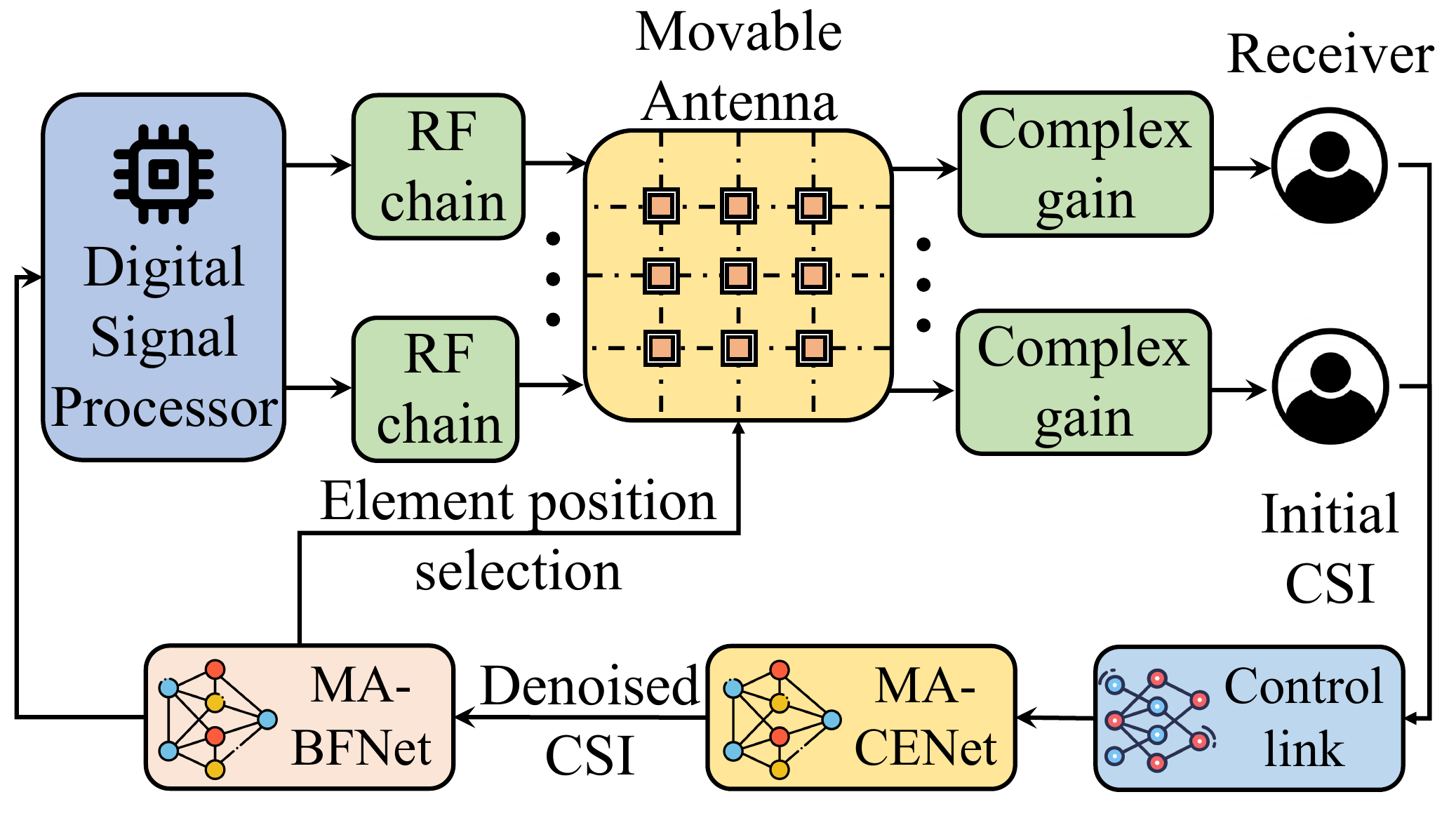}
%\vspace{-1em}
\caption{MA-assisted multiuser communication systems \cite{feng2026deep}}.
\label{Section_3_Figure1}
\vspace{-2em}
\end{figure}

To reduce channel-estimation overhead by learning MA position-control decisions from partial or implicit CSI, a DNN-based method is proposed for MISO systems in \cite{lu2025learning}, where only the measured channel power gains at $K$ sampling points are used to predict the optimized MA-position indices. The adopted up-down MLP first extracts high-level features from partial power measurements and then generates an $N$-dimensional MA-position output. It is trained offline with graph-based optimal MA positions as labels and is used online to determine MA positions without full channel-map estimation. By contrast, the authors in \cite{feng2026deep} develop a complete CE-to-beamforming pipeline for wideband multiuser MA systems. As shown in Fig. \ref{Section_3_Figure1}, limited-pilot CS is first used for initial CE, followed by Swin-Transformer-based MA-CENet for CSI denoising. The refined CSI is then fed into MA-BFNet, where MA-PSN predicts position-selection probabilities, and MA-DBN generates digital beamforming matrices using Transformer encoding and a model-driven WMMSE structure. Simulations show an NMSE gain of about 3 dB over DnCNN and AdaFortiTran, and the highest sum rate among the compared schemes.

To build a channel-estimation-to-beamforming pipeline, the authors in \cite{feng2026deep} propose a DL-based framework for wideband multiuser MA systems. As shown in Fig. \ref{Section_3_Figure1}, limited-pilot CS first provides initial CSI, which is then refined by the Swin Transformer-based MA-CENet using window-based and shifted-window self-attention. The framework is trained with NMSE for CSI denoising and negative sum rate for beamforming optimization. Simulations show about a 3 dB NMSE gain over DnCNN and the Adaptive Fourier Transformer, as well as the highest sum rate among the compared schemes.

In \cite{shao2025hybrid}, a hybrid near-far field six-dimensional MA (6DMA) system is investigated, where a hybrid-field channel model captures planar-wave propagation within each 6DMA surface and spherical-wave propagation among different surfaces. A directional-sparsity-based CE method reconstructs the channel map from limited position-rotation measurements. Based on the estimated channel, 6DMA positions, rotations, and transmit beamforming are jointly optimized by an A2C-based DRL algorithm. The state includes power and user-location information; the action includes 6DMA positions, rotations, and beamforming entries; and the reward is the sum rate with constraint-violation penalties. Simulations show that the proposed CE method outperforms the conventional LS.

%-----------------------------------------------------------------
\subsection{Joint Antenna Position and Sensing}
%-----------------------------------------------------------------

Robust ISAC systems based on cellular and CF structures are first studied under practical synchronization and channel impairments. The authors in \cite{xiu2026robust} propose an MA-aided CF-ISAC architecture to address time synchronization (TS) errors. Then, a Cramer-Rao lower bound (CRLB)-based sensing-accuracy problem is formulated, and a robust framework that combines manifold optimization with MA-enabled meta-RL (MA-MetaRL) is developed. The MA-MetaRL framework combines MAML with TD3, enabling fast policy adaptation across different timing and network conditions. Simulation results show that the proposed method achieves a lower, more stable CRLB than the SCA, DRL, and FPA baselines. The authors in \cite{xiu2025movable} further incorporate imperfect CSI into cooperative ISAC design. A constrained DRL framework is then developed, in which a modified DDPG algorithm with a Wolpertinger architecture is used to handle the mixed, constrained action space. During training, the cumulative reward increases and the cumulative cost decreases, showing that the C-DRL agent learns to reduce power while satisfying communication and sensing constraints. Compared with FPA-based schemes, the MA-aided design achieves lower average transmit power under both CSI and TS errors. Beyond TS errors, \cite{xiu2026meta} studies frequency-domain synchronization impairment in a wideband full-duplex CF-DFRC system with MAs under CFO. A two-stage MO-PDD-MRL framework is proposed, where MO/PDD-based optimization addresses the CFO-robust subproblem, and MRL learns MA positions and beamforming policies in dynamic environments. The MRL state includes CSI for AP-user, self-interference, inter-AP interference, and sensing links, while the action includes beamforming, MA positions, receive filters, and power-related variables. Simulations show faster convergence than conventional DRL and more robust WCSR performance under increasing CFO variations.

The above robust designs mainly rely on RL or meta-RL, whereas another line of work directly learns optimization variables through unsupervised neural networks. For example, \cite{wang2025unsupervised} develops an unsupervised DL framework for a near-field MA-enabled downlink ISAC system. The neural network is trained with a custom loss function that directly reflects the optimization objective and constraint violations. The network outputs both beamforming variables and antenna positions, and constraint-aware modules are embedded to maintain feasible MA layouts. Simulation results show that the proposed DL method consistently outperforms benchmark schemes in sum-rate performance under near-field propagation. Practical hardware impairments are further considered in \cite{zhang2026crosstalk}, where antenna crosstalk is introduced into MA-enabled ISAC beamforming. A TD3-based DRL algorithm is adopted to train a crosstalk-resilient flexible beamforming agent. Simulation results show that the crosstalk-resilient MA-TD3 scheme achieves a lower CRB than the FPA+SDR, FPA+TD3, MA+TD3 (without crosstalk awareness), and random beamforming baselines. Full-duplex ISAC introduces another layer of coupling due to self-interference and uplink/downlink coexistence. To address this issue, \cite{amhaz2026meta} studies a full-duplex (FD) ISAC system with MAs, where an FD base station (BS) serves DL and UL users while a receiving BS captures the reflected echo for sensing. A gradient-based meta-learning (GML) method is proposed for large-scale optimization. The GML architecture uses precoding, uplink power, MA position, communication receive beamforming, and sensing receive beamforming networks to update the coupled variables through inner and outer iterations. Simulation results show that a discretized MA benchmark achieves about $89.7\%$ of the continuous-position architecture, while MA-based schemes outperform FPA benchmarks in reflected-echo SINR. Moving from ISAC to integrated sensing, communication, and power transmission (ISCPT), the authors in \cite{xie2025learning} propose a learning-based flexible beamforming method for MA-enabled ISCPT. A heterogeneous MADDPG framework is adopted, in which beamforming agents optimize beamforming vectors for IoT devices and movable-antenna agents optimize MA positions. The agents share environmental information while the reward balances communication reliability, energy harvesting, and sensing accuracy with penalty terms for constraint violations.

Security-oriented MA-ISAC designs exploit antenna mobility to reshape legitimate and eavesdropping channels. In \cite{le2025beamforming}, secure beamforming is studied for MA-aided ISAC under imperfect eavesdropper CSI. A PPO-based DRL algorithm is used to learn the beamforming and MA movement policy, providing better adaptability than one-shot convex optimization under uncertain CSI. Simulation results show that using MAs improves the secrecy rate by about $40\%$ compared with FPA-based schemes. The authors in \cite{xie2025learning} investigate secure transmission for an active RIS-aided ISAC system with discrete MAs. An AO-based method using penalty-SCA and PDD first provides high-quality beamforming and RIS phase-shift solutions, and then employs a hybrid DRL method, namely dueling Double Deep Q-Network (D3QN)-TD3, to handle the mixed action space. The action includes BS beamforming, RIS phase shifts, and transmit/receive MA position-selection matrices, and the reward combines the secrecy-rate objective with penalties for communication and radar constraints.

The study in \cite{gao2025distributed} considers an ELAA-based downlink ISAC system under user roaming. Two schemes, movable subarray selection (MSS) and adjustable antenna selection (ASS), are proposed to improve beamforming flexibility with low complexity. A dimension-reduction-assisted MAPPO algorithm is developed in which each subarray is treated as an agent. During centralized training, the critic networks share observations among agents to promote cooperation, whereas decentralized execution allows each subarray agent to control its own antenna selection and beamforming.

%-----------------------------------------------------------------
\subsection{Joint Antenna Position and Trajectory/Phase Shift}
%-----------------------------------------------------------------

\begin{figure}[t]
%\vspace{-1em}
\centering
\includegraphics[width=\linewidth]{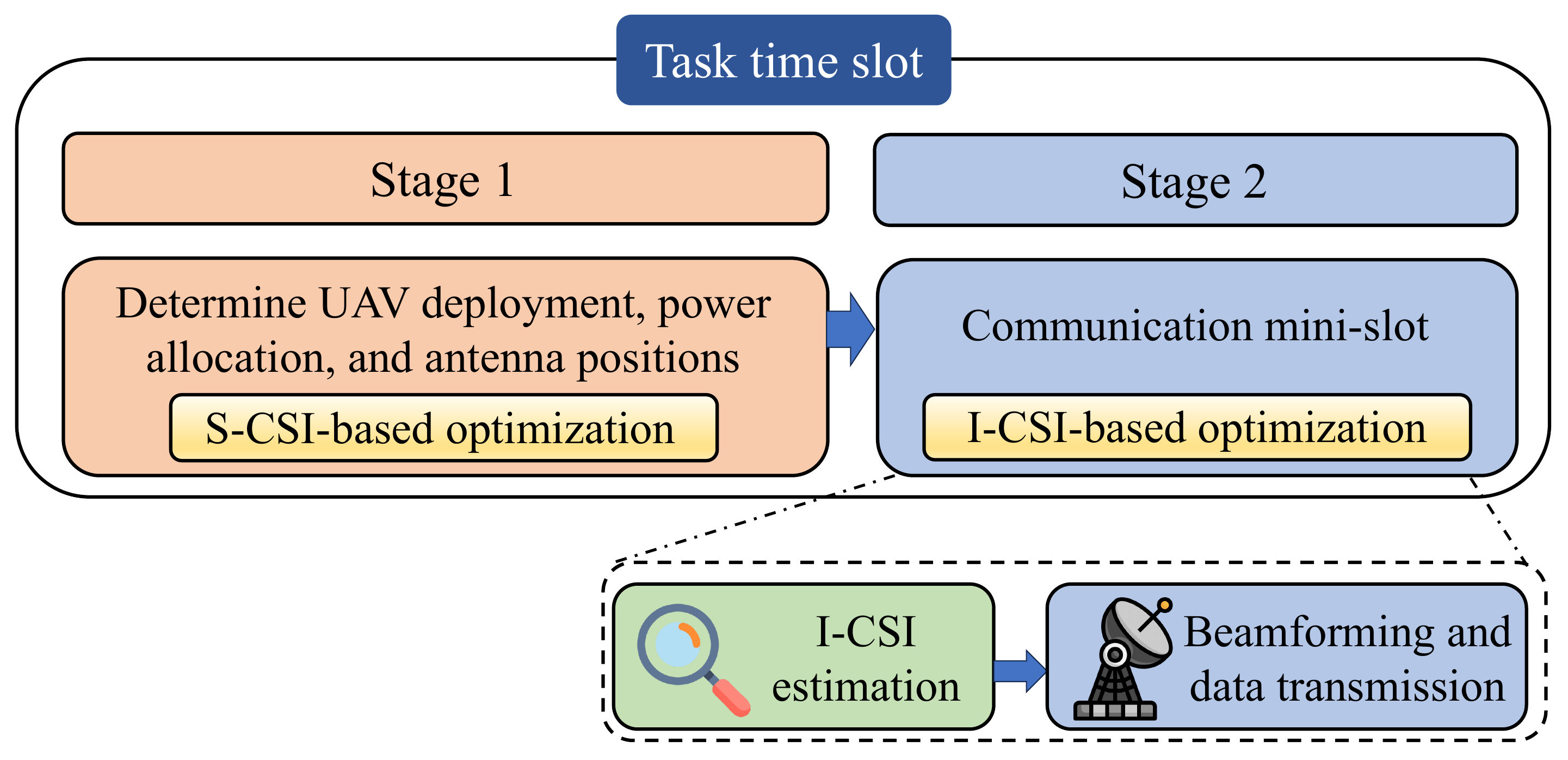}
%\vspace{-1em}
\caption{Two-stage optimization scheme for MA-aided A2G communication systems \cite{zhou2026joint}.}
\label{Section_3_Figure2}
\vspace{-2em}
\end{figure}

\subsubsection{UAV-assisted Systems}
% 4 papers: MA-aided UAV joint deployment, power control and
% beamforming (DRL two-stage, A2G), MA-equipped UAV for backscatter
% data collection (DRL), MA design for UAV-aided federated learning (DRL).

%\cite{zhou2026joint,zhou2025joint,bai2025movable,ahmadzadeh2024movable}

In \cite{ahmadzadeh2024movable}, a UAV-aided over-the-air FL framework is considered, where UAVs act as FL clients and MAs are deployed at the access point. AI plays a central role in this work, as the non-convex dynamic optimization problem is reformulated as a Markov decision process (MDP) and solved by TD3. This enables adaptive antenna and beamforming control in the presence of time-varying UAV locations. Simulation results show that TD3 with movable antennas (MAs) achieves the highest average reward; for example, at 30 users, TD3-MA reaches about $-7$, compared with about $-10$ for SAC-MA and about $-13.5$ for TD3 with FPAs. Moving to communication-oriented trajectory design, an MA-aided air-to-ground (A2G) communication system is investigated in \cite{zhou2025joint}, in which UAV trajectories, antenna positions, and beamforming vectors are jointly optimized to maximize the communication sum rate. A two-stage transmission scheme is developed to reduce the overhead. Specifically, location-aware statistical CSI (S-CSI) is first used to train UAV trajectory and antenna deployment policies, and limited I-CSI is later acquired at the optimized positions to refine beamforming. AI is introduced through a dual space-scale DRL framework, where the outer-layer network learns large-scale UAV trajectory control, and the inner-layer network learns small-scale antenna adjustment. This hierarchical learning structure improves training efficiency by decoupling strongly coupled variables.

As shown in Fig. \ref{Section_3_Figure2}, a more comprehensive A2G design is further studied in \cite{zhou2026joint}, where UAV deployment, MA positions, power control, and beamforming are jointly optimized to maximize downlink sum rate. This work considers a richer set of coupled variables and proposes a dual-spatial-scale hierarchical DRL framework. The outer DRL module learns UAV deployment and power control at the macro spatial scale, while the inner DRL module learns MA position policies at the micro spatial scale. AI is therefore used not only for trajectory-related decision-making, but also for hierarchical variable decoupling and long-term policy learning under reduced CSI overhead. Simulation results show that the proposed dual-spatial-scale hierarchical deep reinforcement learning (DSSH-DRL) scheme achieves clear sum-rate gains over FPA systems; for example, when the antenna span increases from $1.5\lambda \times 1.5\lambda$ to $3\lambda \times 3\lambda$, its gain over FPA expands from about $4$ bps/Hz to $6.5$ bps/Hz, and reaches about 25\% when the span is $4.5\lambda \times 4.5\lambda$. Beyond downlink communication, an MA-equipped UAV for data collection in backscatter sensor networks is considered in \cite{bai2025movable}. The UAV acts as both the RF carrier transmitter and the receiver of backscattered signals, while the directional MA adjusts its orientation to improve the channel gain toward each backscatter device. AI is introduced via a SAC-based DRL strategy, in which the agent uses the azimuth angle and distance between the UAV and the sensors to reduce the observation dimension and stabilize training.

\subsubsection{RIS-assisted Systems}
% 7 papers: MARL for MA + RIS joint optimization, STARS-aided MA
% (learning-based), aerial RIS-aided anti-jamming with movable elements,
% MA-assisted covert communications with RIS, MA SWIPT with STAR-RIS
% (meta DRL), weighted sum-rate with joint MA grouping + movable RIS,
% MA + STARS-RIS robust vehicular networks under imperfect CSI.
% CORRECTION: zhang2026movable moved here from MA-E
% (MA + STARS-RIS paper, not security/resource).

%\cite{zhuang2025multi,yu2025learning,geng2025aerial,xie2024movable,
%amiri2025movable,tang2025weighted,zhang2026movable}

In \cite{zhuang2025multi}, a multi-user MISO system assisted by MAs and RIS is considered, where MA positions, BS beamforming, and RIS phase shifts are jointly optimized to maximize the sum spectral efficiency. A heterogeneous MARL framework is adopted, in which multiple agents optimize MA positions, while another agent controls BS beamforming and RIS phase shifts. AI is therefore used to coordinate highly coupled active and passive beamforming variables in a continuous decision space. Simulation results show that the proposed enhanced TD3 algorithm achieves higher achievable rates than standard TD3 and DDPG. When the number of STAR-RIS elements increases from $N=9$ to $N=36$ with $L=4$ transmit paths, the achievable rate improves by about 40\%. To overcome the half-space limitation of conventional RIS, a STAR-RIS-aided communication system with MAs is investigated in \cite{yu2025learning}. The MA coordinates, transmit beamforming, and STAR-RIS transmission/reflection coefficients are jointly optimized to maximize the weighted sum rate. A TD3-based learning framework is introduced to search for feasible control policies subject to coupled constraints, with penalty mechanisms incorporated to regulate action outputs.

The integration of STAR-RIS, MAs, and wireless power transfer is further explored in \cite{amiri2025movable}, where a STAR-RIS-assisted MA SWIPT system is developed for multi-user downlink transmission. A meta-TD3-based method is adopted to improve adaptation across different system configurations. In this case, AI serves not only as a resource allocation tool but also as a fast adaptation mechanism for dynamic SWIPT scenarios. Simulation results show that meta TD3 (MTD3) outperforms conventional TD3; for example, at an average data rate of about $3.5$ bps/Hz, MTD3 achieves around $2.3$ average harvested energy, compared with about $1.8$ for TD3, while also maintaining better convergence stability. Moving from rate and energy optimization to security-oriented design, covert communication with MAs and RIS is studied in \cite{xie2024movable}. The MA trajectories, transmit beamforming, and RIS phase shifts are jointly optimized to improve the covert rate. The MA movement trajectory is modeled as an MDP and optimized by DRL, while beamforming and RIS phase shifts are handled through AO. AI is therefore primarily responsible for learning long-term spatial movement policies that increase channel uncertainty, thereby avoiding detection.

Anti-jamming transmission with aerial RIS and movable elements represents another important extension of RIS-assisted systems. In \cite{geng2025aerial}, an aerial RIS-assisted anti-jamming system with movable elements is considered, where the ARIS trajectory, passive beamforming, and BS active beamforming are jointly optimized to maximize the system sum rate. A DDQN-based method is adopted to address the non-convex and dynamically complex decision problem. AI is used to learn anti-jamming transmission strategies under uncertain interference conditions, where movable RIS elements provide additional spatial flexibility compared with fixed-element ARIS. Simulation results show that a movable-element aerial reconfigurable intelligent surface with double deep Q-network (ME-ARIS DDQN) achieves the highest achievable sum rate; for example, after convergence, it reaches about $41$ bps/Hz, compared with about $39$ bps/Hz for a fixed-position aerial reconfigurable intelligent surface with DDQN (FP-ARIS DDQN), about $38$ bps/Hz for ME-ARIS with DQN, and about $24$ bps/Hz for the random baseline. In addition to aerial RIS mobility, RIS location optimization is studied from a more infrastructure-oriented perspective in \cite{tang2025weighted}. A movable RIS is mounted on a slide rail, and BS transmit beamforming, RIS passive beamforming, and RIS location are jointly optimized to maximize the weighted sum rate. Product-manifold optimization is first used to handle the coupled beamforming and location variables, while an enhanced gradient-based manifold meta-learning method is further introduced for large-scale scenarios. In this work, AI is used to accelerate adaptation and reduce complexity in optimizing large-scale RIS deployment.

Robustness under imperfect CSI is also relevant to RIS-assisted and movable reconfigurable systems. In \cite{zhang2026movable}, an MA-enabled vehicular consumer network is considered under imperfect CSI, in which MA positions and transmit beamforming are jointly optimized to maximize the worst-case sum rate. Specifically, a robust AO framework is developed by using deterministic reformulation, the S-Procedure, the Schur complement, and successive convex approximation. This work provides an optimization-based robustness perspective for high-mobility environments, which complements AI-driven RIS/MA designs by emphasizing worst-case performance guarantees. Simulation results show that the proposed robust MA scheme achieves near-upper-bound max-min rate and clearly outperforms FPA and FB baselines; for example, when the number of antennas increases to $8$, it reaches about $8.3$ bps/Hz, compared with about $6$ bps/Hz for FPA and about $5.2$ bps/Hz for FB.

%-----------------------------------------------------------------
\subsection{Joint Antenna Position and Security/Resource Allocation}
%-----------------------------------------------------------------
% 10 papers: MA for URLLC (DRL), MA relay position optimization (RL),
% C-NOMA with MA (DDPG), MA uplink RSMA for URLLC, Transformer for
% MA position prediction for secure comms, deep learning jamming
% mitigation with MA, deep unfolding jamming mitigation, MA array
% jamming mitigation, CoMP-RSMA with MA (meta-learning),
% MA vs UAV for PLS.
% CORRECTION: zhang2026movable removed (moved to MA-D, RIS paper).

%\cite{liu2026movable,zhang2025reinforcement,amhaz2025optimizing,
%fang2026toward,wang2026moving,tang2025deep,zhang2025deep,
%guo2025movable,amhaz2026enhancing,li2025can}

In resource-allocation-oriented designs, MAs are primarily used to improve spectral efficiency, reliability, and service quality under stringent communication constraints. In \cite{liu2026movable}, an MA-enabled uplink URLLC system is considered, where MA positions and users' transmit power are jointly optimized to maximize the weighted sum rate under finite-blocklength transmission. The optimization problem is solved using a TD3-based DRL algorithm, and the AI learns continuous policies for antenna position and power control without relying on highly complex analytical optimization. Simulation results show that the MA-aided RSMA scheme achieves clear sum-rate gains over the FPA-based RSMA scheme; for example, it improves the sum rate by 34.8\% at 0 dB and by 27.4\% at 30 dB, while adaptive decoding further improves rate-splitting multiple access with fixed decoding order (RSMA-FDO) by 2.9\% at $K=4$ and 3.4\% at $K=6$. To enhance spectral efficiency for reliable low-latency services, an MA-aided uplink RSMA framework is investigated in \cite{fang2026toward}, in which antenna positioning, receive combining, power allocation, and decoding order are jointly optimized for mobile broadband reliable low-latency communication (mBRLLC) services. A tree search algorithm with channel-aware dynamic pruning is combined with heterogeneous MARL to handle the combinatorial SIC decoding order and continuous resource variables. In this case, AI is used to coordinate heterogeneous decision modules and reduce the complexity of mixed-integer resource management. Moving from direct uplink transmission to relay-assisted communication, an MA-enabled buffer-aided relay system is studied in \cite{zhang2025reinforcement}, where the relay MA's position and the transmit powers of the source and relay nodes are dynamically optimized. DDPG is adopted to learn long-term antenna movement and power allocation policies under buffer, power supply, and antenna movement energy constraints.

More advanced resource management frameworks further combine MAs with cooperative transmission and non-orthogonal access. In \cite{amhaz2026enhancing}, a coordinated multi-point (CoMP)-RSMA downlink system with user-side MAs is considered, where BS transmit beamforming, common stream allocation, and MA positions are jointly optimized under QoS constraints. A gradient-based meta-learning algorithm is introduced to handle large-scale optimization without extensive pre-training, with AI primarily serving as a fast adaptation tool to solve the coupled CoMP-RSMA resource allocation problem. Simulation results show that the proposed CoMP-RSMA-MA scheme achieves over 97\% of the fmincon-based near-optimal performance and provides up to an 80\% gain over CoMP-RSMA-FPA and a 180\% gain over CoMP-SDMA-MA at a BS power budget of 37 dBm. In \cite{amhaz2025optimizing}, an MA-enabled downlink cooperative NOMA system is investigated, in which BS beamforming, device transmit power, and MA positions at users are jointly optimized. A DDPG-based method is adopted to handle the continuous state and action spaces arising from user cooperation and antenna mobility, thereby enabling online control of cooperative transmission and MA placement.

Security-oriented studies focus on exploiting antenna mobility to reshape legitimate and illegitimate channels. In \cite{wang2026moving}, the MA positioning problem for secure wireless communications is reformulated as a prediction task rather than a repeated real-time optimization problem. A role-aware Transformer framework, namely RoleAware-MAPP, is developed by incorporating role-aware embeddings, physics-informed semantic features, and a secrecy-prioritized loss function. Specifically, Transformer-based learning can be used to map CSI and user-location features into future MA positions, thereby replacing repeated online optimization with data-driven position prediction. Simulation results show that RoleAware-MAPP achieves an average secrecy rate of $0.3606$ bps/Hz, improving over the Transformer baseline by 35.5\% and DDPG by 13.4\%, while also increasing SPSC over the Transformer baseline by 6.73 percentage points. From a broader physical-layer security perspective, MA-enabled micro-mobility and UAV-enabled macro-mobility are compared in \cite{li2025can}. The average secrecy rate is maximized by jointly optimizing MA positions and beamforming for the micro-mobility scheme, and UAV trajectory and beamforming for the macro-mobility scheme. Although this work is mainly optimization-driven rather than AI-driven, it provides an important security-oriented benchmark for understanding when local antenna movement can replace or complement UAV trajectory control.

Anti-jamming studies further show the value of MAs in adversarial wireless environments. In \cite{tang2025deep}, a receiver-side MA array is considered for communication in the presence of multiple jammers. The receive beamforming and antenna element positions are jointly optimized to maximize SINR, with beamforming handled via a Rayleigh-quotient formulation and antenna positioning learned by a multi-layer perceptron (MLP). AI is used to replace repeated online antenna-position optimization with offline training and low-complexity inference. Simulation results show that the proposed DL-assisted MA scheme achieves near-AO SINR with much lower runtime; for example, with $8$ antenna elements, it reaches about $6.5$ SINR, close to AO at about $6.7$, while clearly outperforming FPA at about $3.7$ and RPB at about $2.5$. To improve robustness to imperfect jamming CSI, an MA array-based jamming mitigation scheme is developed in \cite{guo2025movable}, where receive beamforming and the antenna position vector are jointly optimized under angular uncertainty in the jamming. An AO framework with stochastic successive convex approximation is adopted, and a model-driven DL scheme is further introduced to accelerate convergence. In this work, AI primarily assists the optimization process rather than fully replacing it, thereby improving efficiency while retaining the model structure. To further improve interpretability and reduce computational overhead, a deep unfolding-based anti-jamming framework is proposed in \cite{zhang2025deep}. The non-convex joint optimization of receive beamforming and antenna positions is first decoupled, and the iterative optimization steps are then unfolded into neural network layers. AI is therefore embedded into the optimization procedure in a model-driven manner, enabling faster convergence and better interpretability than black-box learning. Simulation results show that the proposed deep unfolded jamming mitigation (DUJM) scheme converges to a near-maximum SINR of about $30$ within roughly $35$ epochs and achieves the strongest main-lobe gain with effective jamming nulls, outperforming MA, FPA, and antenna-selection baselines in beam-pattern-based interference suppression.

\subsection{Lessons Learned}

In this Section, the reviewed studies show that AI has become an effective tool for exploiting the spatial flexibility of MAs in beamforming, channel estimation, sensing, trajectory/phase-shift design, and security/resource allocation. Its main roles are to reduce the complexity of online optimization, improve adaptation to dynamic environments, and coordinate strongly coupled design variables. Supervised learning and Transformers are mainly used for fast prediction; DRL and MARL for sequential decision-making; meta-learning for rapid adaptation; and deep unfolding for interpretable, model-driven learning. However, practical deployment remains limited by simplified channel models, fixed system dimensions, idealized CSI assumptions, and offline training distributions.

%=====================================================================
\section{AI for Fluid Antenna Systems}
\label{sec:ai_for_FAs}
%=====================================================================
% Overview paragraph introducing the section and the 5 subsections
%{\bf{{\color{red}We need a paragraph here to summarize this section, just like the first paragraph of Section~III.}}}

This section reviews AI-assisted port selection and joint optimization techniques in FASs, in which antenna ports can be dynamically activated within a compact spatial region to exploit spatial diversity and interference fluctuations. Compared with MA systems, FASs usually involve a large number of discrete candidate ports, making port observation, port-state inference, dynamic switching, and resource coordination highly coupled and computationally challenging. Therefore, AI techniques, including supervised learning, DRL, MARL, Transformers, GNNs, LLMs, and deep unfolding, have been increasingly employed to learn efficient port-selection policies, infer unobserved channel states, reduce port-search overhead, and coordinate beamforming, sensing, trajectory design, resource allocation, and computation offloading.
%-----------------------------------------------------------------

%-----------------------------------------------------------------
\subsection{Joint Port Selection and User Admission}
%-----------------------------------------------------------------
%User Scheduling and Port Selection: \cite{waqar2024opportunistic,gu2025partially,li2026ai}

%Low-Overhead FAMA: \cite{fan2025ai,waqar2023deep,eskandari2024cgan,waqar2026attentional}

%Port Prediction: \cite{zhang2023fast,wong2024virtual,silveira2025comparison}

%Dynamic Port Switching: \cite{liu2026switching,peng2026group}

%\cite{waqar2024opportunistic}\\
\begin{figure}
    \centering
    \includegraphics[width=0.9\linewidth]{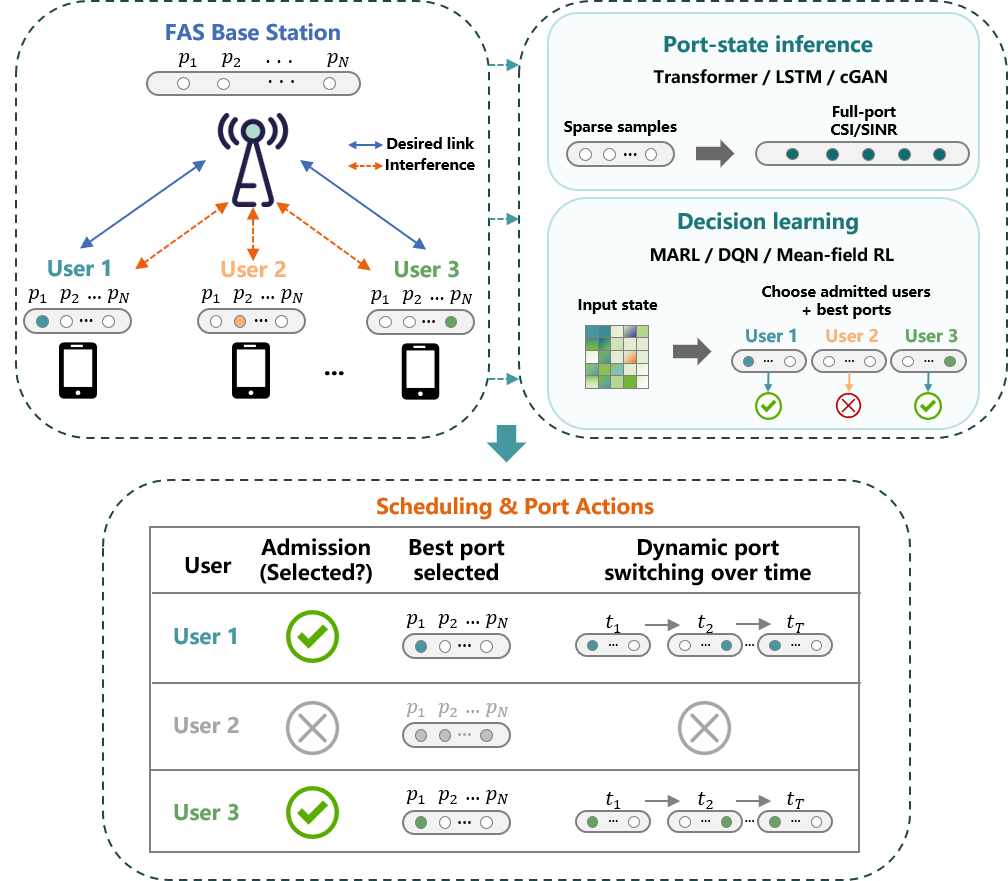}
    \caption{AI for Joint Port Selection and User Admission in Fluid Antenna Systems.}
    \label{fig:section4_user_admission}
    \vspace{-2em}
\end{figure}

\begin{table*}[t]
\centering
\caption{Summary of AI for Fluid Antenna Systems: Joint Port Selection and User Admission}
\label{tab:ai_fas_part1}
\renewcommand{\arraystretch}{1.2}
\begin{tabular}{|l|c|l|l|l|l|}
\hline
\rowcolor{myblue}
\textbf{Learning Stage} & \textbf{Ref.} & \textbf{Port Observation} & \textbf{Optimization Target} & \textbf{Method} & \textbf{Port Decision} \\
\hline

\multirow{3}{*}{\makecell[l]{Admission-aware\\Port Selection}}
& \cite{waqar2024opportunistic}
& O-FAMA channel state
& Robust sum-rate maximization
& Team-inspired MARL
& User-port matching \\
\cline{2-6}

& \cite{gu2025partially}
& Partial SINR observations
& Joint access-port control
& Mean-field RL
& Decentralized access \\
\cline{2-6}

& \cite{li2026ai}
& Partial port CSI and packets
& Long-term AoI minimization
& Transformer + DDQN
& Low-delay scheduling \\
\hline

\multirow{4}{*}{\makecell[l]{Sparse Port\\Observation}}
& \cite{fan2025ai}
& Correlated port fading
& SIR-based port selection
& DQN
& Sequential port search \\
\cline{2-6}

& \cite{waqar2023deep}
& Few observed SINRs
& Port-search overhead reduction
& LSTM
& SINR completion \\
\cline{2-6}

& \cite{eskandari2024cgan}
& Limited SINR samples
& SINR envelope recovery
& cGAN
& SINR reconstruction \\
\cline{2-6}

& \cite{waqar2026attentional}
& Sparse turbo-FAMA samples
& Full port-field recovery
& Attentional copula
& Correlation extrapolation \\
\hline

\multirow{3}{*}{\makecell[l]{Port Prediction\\and Virtual Ports}}
& \cite{zhang2023fast}
& Sparse CSI over time
& All-port CSI prediction
& LSTM
& Fast CSI prediction \\
\cline{2-6}

& \cite{wong2024virtual}
& Physical-port signals
& Virtual signal inference
& Transformer + RZF
& Virtual-port inference \\
\cline{2-6}

& \cite{silveira2025comparison}
& Spatially correlated channels
& Outage probability reduction
& Classifier / regressor
& Best-port prediction \\
\hline

\multirow{2}{*}{\makecell[l]{Dynamic Switching\\and Robustness}}
& \cite{liu2026switching}
& Spatio-temporal channels
& Throughput-cost tradeoff
& Transformer-DDQN
& Cost-aware switching \\
\cline{2-6}

& \cite{peng2026group}
& Imperfect CSI
& Robust position optimization
& GRPO
& Robust switching \\
\hline

\end{tabular}
%\vspace{-2em}
\end{table*}

AI provides an effective tool for joint user admission and port selection in FASs, where admission control, sparse port observation, port-state inference, and dynamic switching are tightly coupled. Fig.~\ref{fig:section4_user_admission} illustrates the overall AI-assisted decision pipeline, while Table~\ref{tab:ai_fas_part1} summarizes representative studies from the perspectives of learning stage, optimization objective, and port-decision strategy.

Opportunistic FAMA naturally couples user scheduling with port selection as the system must determine not only which users are admitted for transmission but also which FAS ports to activate for those users. In \cite{waqar2024opportunistic}, opportunistic user scheduling and port selection are jointly investigated for O-FAMA systems, in which favorable users are selected for transmission, and each selected user selects its best FAS port to suppress multi-user interference. A team-inspired multi-agent RL framework is developed to solve the robust sum-rate maximization problem in a decentralized manner, with derivative-assisted learning used to improve convergence. The authors in \cite{gu2025partially} further study joint scheduling and port selection under partial observability, where each UE independently decides whether to access the channel and which port to select based on limited SINR observations. To reduce the complexity of joint action optimization, mean-field RL is used to approximate the impact of other UEs by modeling their average behavior.

%\cite{gu2025partially,li2026ai,fan2025ai,waqar2023deep,eskandari2024cgan}\\

For short-packet next-generation ultra-reliable and low-latency communications (xURLLC) systems, the work in \cite{li2026ai} minimizes the long-term average age of information (AoI) by jointly optimizing user scheduling and port selection under partial port CSI and stochastic packet arrivals. A mask-guided port-embedding Transformer reconstructs full-port CSI, while a curriculum-learning-based dueling double DQN optimizes user scheduling and port selection. Near-zero CSI reconstruction error is achieved with 10\% observed ports, while curriculum learning improves AoI by 38.38\%--40.75\% at a 5\% observation ratio.

Another line of research focuses on reducing the port observation and search overhead in FAMA. The authors in \cite{fan2025ai} formulate adaptive port selection as a sequential decision-making problem, where a DQN agent selects the port that maximizes the signal-to-interference ratio (SIR) under correlated fading and multi-user interference. The work in \cite{waqar2023deep} focuses on slow FAMA, where an LSTM-based model infers the SINRs of unobserved ports from a small number of observations, thereby reducing port-search overhead. Generative and extrapolation-based methods are also explored for this purpose. The authors in \cite{eskandari2024cgan} employ a cGAN to generate the SINR envelope of unobserved ports from limited SINR samples, enabling port selection without exhaustive scanning. The work in \cite{waqar2026attentional} extends this idea to turbo FAMA, where a two-stage attentional-copula extrapolator reconstructs the full port field from sparse observations by modeling both channel-signal coupling and inter-port spatial correlation.

%\cite{waqar2026attentional}\\

Apart from FAMA-specific access designs, general FAS port prediction and virtual-port learning have also been explored. The work in \cite{zhang2023fast} exploits spatial and temporal correlations for fast port selection, in which an LSTM-based method estimates all-port CSI from sparse observations and predicts CSI in later time slots. The authors in \cite{wong2024virtual} propose a virtual FAS architecture based on learning-based imaginary antennas, in which a Transformer-based predictor infers signals at virtual positions and improves interference handling using regularized zero-forcing (RZF) detection. In \cite{silveira2025comparison}, multi-label classification and regression models are compared for best-port prediction, showing that both methods benefit from spatial correlation, with multi-label classification achieving slightly better outage performance.

%\cite{zhang2023fast,wong2024virtual,silveira2025comparison,peng2026group,ho2025deep,liu2026switching}\\

Practical FAS deployment further requires that port-switching decisions consider temporal dynamics, switching costs, and robustness. The work in \cite{liu2026switching} studies switching-cost-aware dynamic port selection, where long-term throughput and physical switching cost are balanced under spatio-temporally correlated channels. A Transformer-assisted dueling DQN with candidate pre-selection and local refinement is developed to achieve near-optimal performance with much lower latency than exhaustive search. Robust switching-position optimization is further investigated in \cite{peng2026group} for FAS-driven blind interference alignment, where group-relative policy optimization (GRPO) is used to optimize switching positions under imperfect CSI and to improve robustness to poor local optima.

%\cite{ho2025deep}: overview

\subsection{Joint Port Selection and Beamforming Optimization}
%-----------------------------------------------------------------

%In this subsection, there may have some papers that are only port selection without beamforming optimization.
%If so, you can move them in Subsection A. 

\begin{table*}[t]
\centering
\caption{Summary of AI for Fluid Antenna Systems: Joint Port Selection and Beamforming Optimization}
\label{tab:ai_fas_part2}
%\scriptsize
\renewcommand{\arraystretch}{1.2}
\begin{tabular}{|l|c|l|l|l|l|}
\hline
\rowcolor{myblue}
\textbf{Learning Stage} & \textbf{Ref.} & \textbf{Port Observation} & \textbf{Optimization Target} & \textbf{Method} & \textbf{Port Decision} \\
\hline

\multirow{3}{*}{\makecell[l]{Position--BF\\Co-Design}}
& \cite{fan2025spectral}
& MU-MISO FAS
& SE maximization
& Averaged-DQN
& FA position control \\
\cline{2-6}

& \cite{zhang2025indoor}
& Indoor FAS
& Layout-aware sum-rate max.
& GRPO
& Geometry-aware positioning \\
\cline{2-6}

& \cite{kharouaa2025ai}
& Multi-user FAS
& SE optimization
& CNN + DRL
& Hybrid BF/position control \\
\hline

\multirow{3}{*}{\makecell[l]{LLM-Assisted\\Joint Design}}
& \cite{guo2025llm}
& MISO-FAS
& Sum-rate maximization
& LLM + LoRA
& Port/BF generation \\
\cline{2-6}

& \cite{wang2025large}
& General FAS
& Port-precoder design
& LLM framework
& Reasoning-guided design \\
\cline{2-6}

& \cite{wang2025llm}
& FAS precoding
& Joint port-precoder opt.
& AO + dual LLM
& Hyper-heuristic search \\
\hline

\multirow{2}{*}{\makecell[l]{Graph Learning\\\& Hardware Design}}
& \cite{xu2025toward}
& FPGA FAS
& Low-latency deployment
& GNN-RPS + FPGA
& Fast BF inference \\
\cline{2-6}

& \cite{he2025graph}
& MU-MISO FAS
& Position-BF optimization
& Two-stage GNN
& Graph-based coupling \\
\hline

\multirow{3}{*}{\makecell[l]{RIS/BDRIS/FRIS-\\Assisted FAS}}
& \cite{chen2025unified}
& RIS/STAR-RIS FAS
& Joint surface-FA design
& RZF + CEO
& Surface/FA coordination \\
\cline{2-6}

& \cite{anjum2025service}
& BDRIS-FAS
& Service fairness max.
& PPO
& Fairness-aware tuning \\
\cline{2-6}

& \cite{vega2026exploring}
& Fluid RIS
& Secrecy outage reduction
& MLE + Q-learning
& Secure reconfiguration \\
\hline

\multirow{4}{*}{\makecell[l]{Task-Driven\\FAS}}
& \cite{ahmadzadeh2025ai}
& OTA-FL FAS
& Selected-user max.
& LSTM-DDPG
& FL-oriented selection \\
\cline{2-6}

& \cite{ahmadzadeh2025enhanced}
& OTA-FL FAS
& Optimality-gap minimization
& RDPG
& Dynamic FL control \\
\cline{2-6}

& \cite{zhang2025energy}
& IDET FAS
& Energy efficiency max.
& AO + C-SAC
& Energy-aware switching \\
\cline{2-6}

& \cite{ju2026joint}
& MEC FAS
& System-delay minimization
& IB-CE + HiTDMA
& Cross-layer coordination \\
\hline

\multirow{3}{*}{\makecell[l]{Large-Scale\\Networks}}
& \cite{li2025model}
& Multi-cell FAS
& Distributed joint design
& Model-based MARL
& Multi-agent coordination \\
\cline{2-6}

& \cite{zhang2025fars}
& FAS-RSMA NTN
& User fairness max.
& Mamba + transfer learning
& Sequence-aware optimization \\
\cline{2-6}

& \cite{pakravan2025fluid}
& NOMA-FAS
& Sum-rate maximization
& DRL
& Interference-aware control \\
\hline

\end{tabular}
\vspace{-2em}
\end{table*}

Joint port selection and beamforming optimization are central design problems in FASs as the effective channel depends not only on propagation conditions but also on the selected ports or antenna positions. Therefore, the transmit beamforming or precoding vectors have to be optimized jointly with the spatial configuration of the fluid antennas. Such a coupling usually results in a mixed discrete--continuous, highly non-convex problem, in which port indices or antenna positions determine the channel matrix, while beamforming variables further affect interference, transmit power, user fairness, and system throughput. To address this challenge, recent studies have increasingly adopted AI-assisted methods, including DRL, GNNs, LLMs, and sequence models.

Spectral efficiency maximization is investigated for a downlink Multi-user multiple-input single-output (MU-MISO) fluid antenna system by jointly considering antenna positioning and linear precoding in \cite{fan2025spectral}. The problem is formulated under transmit-power, minimum inter-antenna distance, and bounded deployment-region constraints. An Averaged-DQN framework is developed to sequentially adjust antenna positions while updating the corresponding precoding vectors through a deterministic precoding method. Attention-enhanced state encoding, multiple target networks, and a dueling Q-network architecture are further incorporated to improve learning stability and generalization. In \cite{zhang2025indoor}, indoor FASs are studied by jointly optimizing antenna positioning, beamforming, and power allocation under a layout-aware channel model. By capturing LoS and NLoS components induced by the physical layout, this study shows that AI-based joint design can be strongly environment-dependent, since antenna positioning and beamforming are optimized over channels shaped by walls, reflections, and indoor geometry. The authors in \cite{kharouaa2025ai} propose an AI-enabled spectral-efficient fluid antenna system (AI-SEFAS) framework for spectral-efficiency maximization, where CNNs are employed for fast beamforming-weight prediction, and DRL is used for adaptive antenna repositioning. This framework integrates learning-based beamforming and position control to cope with user mobility, channel variation, and multi-user interference.

LLM-assisted solutions have also been explored for joint port-selection and beamforming design. In \cite{guo2025llm}, sum-rate maximization is formulated for multiuser MISO-FAS under transmit-power and port-activation constraints. A parallel-output LLM framework is developed to simultaneously predict port indices and beamforming coefficients, while low-rank adaptation (LoRA) fine-tuning and Gumbel--Sinkhorn relaxation are adopted to reduce training cost and enable differentiable port selection. A broader LLM-empowered design vision is presented in \cite{wang2025large}, where joint port selection and precoder design are treated as an NP-hard combinatorial problem, and LLMs are proposed to assist with problem decomposition, heuristic generation, and automatic parameter tuning. Moreover, the authors in \cite{wang2025llm} propose a hyper-heuristic design in which AO decomposes the original problem into a convex precoder subproblem and a port-selection subproblem, and a dual-LLM architecture is used to enhance the genetic algorithm for port selection.

GNN-based methods have been investigated to reduce the complexity of joint optimization. A practical hardware-software co-design framework is developed in \cite{xu2025toward}, in which GNN-based beamforming inference is combined with random port selection and further implemented on an FPGA accelerator. This study shifts the focus from algorithmic performance alone to practical latency and hardware execution efficiency. In \cite{he2025graph}, a two-stage GNN is designed for MU-MISO FASs, where the first stage learns antenna positions and the second stage learns beamforming vectors. By jointly training the two stages, the method captures the dependence between channel reconfiguration and transmit design while maintaining a modular structure.

%\cite{kharouaa2025ai,guo2025llm,wang2025large,chen2025unified, wang2025llm,xu2025toward,anjum2025service,he2025graph}\\
%\cite{ahmadzadeh2025ai}: Joint beamforming\\
%\cite{zhang2025energy}: Joint beamforming\\
%\cite{ahmadzadeh2025enhanced}: Joint beamforming\\
%\cite{ju2026joint}: Port sselection + beamforming\\
%\cite{vega2026exploring}: Joint with IRS beamforming\\

Beyond standard downlink FAS settings, joint port/position and beamforming design has also been studied in RIS-assisted and task-oriented systems. In \cite{chen2025unified}, transmit beamforming, discrete FA positions, and low-resolution RIS/STAR-RIS coefficients are jointly optimized for MU-MISO downlink systems through RZF precoding and a cross-entropy-based probabilistic learning method. The authors in \cite{anjum2025service} consider BDRIS-assisted FASs and jointly optimize the antenna position vector, beamforming vectors, and BDRIS configuration matrix to improve service fairness under a power budget. In a related physical-layer security scenario, fluid reconfigurable intelligent surfaces are investigated in \cite{vega2026exploring}, where Q-learning is used for selecting spatial element positions, and beamforming/phase-shift design are jointly optimized to enhance secrecy performance. These studies show that the beamforming objective can be extended beyond the base-station precoder to include RIS/BDRIS/FRIS coefficients, further strengthening the coupling among spatial reconfiguration, beamforming, and surface control.

Task-driven FAS optimization has also attracted considerable attention. The integration of FAs into over-the-air FL is studied in \cite{ahmadzadeh2025ai}, where FA positions, the receive beamforming vector, and user selection are jointly optimized to maximize the number of selected users under a mean-squared-error constraint. The work in \cite{ahmadzadeh2025enhanced} minimizes the optimality gap of OTA-FL by jointly optimizing FA positions, the beamforming vector, and user transmit-power allocation with an RDPG-based solver. For integrated data and energy transfer, the authors in \cite{zhang2025energy} consider both switching delay and switching energy consumption caused by port selection, and combine AO for short-term beamforming/port-selection design with constrained SAC for long-term policy learning. Furthermore, FA-assisted MEC networks are investigated in \cite{ju2026joint}, in which channel estimation, port selection, beamforming, power control, and resource allocation are jointly considered to minimize delay.

%\cite{li2025model}: joint precoding\\
%\cite{zhang2025fars}: Joint beamforming\\
%\cite{pakravan2025fluid}: Joint beamforming\\

Another important research direction is extending joint port/position and beamforming optimization to more complex network architectures, such as multi-cell, NTN, and NOMA systems. A model-based multi-agent RL framework is proposed in \cite{li2025model} for multi-cell FASs, where each BS acts as an independent agent and jointly optimizes port selection and precoding in a distributed manner. In \cite{zhang2025fars}, FAS is combined with RSMA in non-terrestrial networks, and antenna port locations, precoder design, and common-rate allocation are jointly optimized for user fairness through a Mamba-based learning solver. Uplink NOMA with imperfect SIC is studied in \cite{pakravan2025fluid}, where FA positions, the BS beamforming vector, and user transmit powers are jointly optimized using a DRL framework. Overall, these studies show that AI-enabled joint port selection and beamforming optimization is evolving from single-link rate improvement toward large-scale, multi-objective, and cross-layer FAS design.

%-----------------------------------------------------------------
\subsection{Joint Port Selection and Channel Estimation}
%-----------------------------------------------------------------
% \cite{yang2025fas,gao2025ssnet,dos2025dl,jiang2025beam,
% zhang2024learning,zhao2026taaformer,recy2026topology}
FAS provides unprecedented spatial DoF by dynamically switching among densely packed antenna ports, but conventional full-port CSI acquisition through pilots incurs prohibitive overhead, complexity, and latency. Moreover, interpolation and compressive sensing are inadequate for capturing the highly non-linear and dense multipath components (MPC) of FAS channels. Consequently, AI has emerged as a pivotal solution, enabling full-port CSI extrapolation from sparse observations or direct port selection without explicit full CSI recovery, which effectively addresses the limitations of traditional methods.

FAS channel estimation can be modeled as a graph or image completion problem by leveraging the inherent spatial correlation across densely packed ports. The authors in \cite{zhang2024learning} propose an asymmetric graph masked autoencoder (AGMAE) framework for spatial-channel extrapolation. By modeling the FAS ports as a graph, AGMAE utilizes a Transformer-based encoder to construct basis vectors and a graph attention network (GAT) decoder to capture local correlations and smoothness. With this local diffusion mechanism, the network can accurately reconstruct the full CSI using only 5\%-10\% of the observed antenna ports. However, this framework relies on large labeled datasets for offline training and is sensitive to noise in low-SNR regimes. To reduce data dependence and enhance robustness against noise and hardware imperfections, a self-supervised learning network (SSNet) is introduced in \cite{gao2025ssnet}. Formulating FAS channel extrapolation as an image reconstruction problem, SSNet integrates a mixture-of-experts (MoE) module to learn generalized representations directly from raw CSI. SSNet enables high-precision FAS channel extrapolation with only 5\%-10\% of the observable ports and achieves remarkable performance when trained on only 2.5\% of the dataset. Meanwhile, the self-supervised nature of this approach effectively reduces labeling cost and offers strong generalization. However, the attention modules and MoE routing introduce substantial inference complexity and latency.

While existing spatial reconstruction models reduce pilot overhead, conventional self-attention suffers from quadratic complexity, limiting its scalability to large-scale FAS. To address this issue, the authors in \cite{zhao2026taaformer} propose the transformer-assisted antenna/port estimation network (TAAformer), which is equipped with a novel transposed angular attention (TAA) mechanism. Instead of directly modeling spatial port correlation, TAA disentangles MPC via an angular-domain transformation and employs a transposed operation to model intrinsic relationships among angular features. Built upon a UNet-like encoder-decoder structure, TAAformer enables multi-scale angular feature extraction, reducing complexity to linear order while retaining global modeling capabilities. It can effectively capture critical MPC features even with low embedding dimensions, delivering superior channel estimation accuracy compared to conventional self-attention methods. However, the TAAformer primarily operates in static environments and lacks temporal modeling to track rapid channel variations.
Highly dynamic FAS scenarios, such as vehicular networks and LEO satellite communications, suffer from rapid channel fluctuations that quickly render instantaneous CSI outdated. Conventional methods only adapt to static or quasi-static environments, which necessitates reliable temporal sequence modeling. Against IEEE 802.11p vehicular networks, a DL-aided data-pilot-assisted (DPA) channel estimation scheme is proposed in \cite{dos2025dl}. Leveraging an LSTM network to capture temporal channel correlations, it predicts the SNR of all FAS ports from only a small set of observed ports, effectively suppressing error propagation under high mobility. Meanwhile, FAS delivers at least an 8 dB performance gain over single-antenna systems, and the two-port combination provides an additional gain of over 15 dB. However, LSTM-based models lack sufficient representational capacity to cope with high-dimensional variations, including delay and Doppler shifts. Extending to high-mobility LEO satellite-FAS scenarios, \cite{yang2025fas} introduces FAS-LLM, a channel prediction architecture for OTFS-enabled downlinks. FAS-LLM adopts a LoRA-tuned LLaMA-3-1B model and combines it with separable principal component analysis (PCA) compression, thereby reducing the input dimensionality by over 99\% and efficiently extracting delay-Doppler channel features. This framework achieves a normalized mean squared error (NMSE) improvement of up to 10 dB in multi-step prediction, enabling proactive port switching in non-terrestrial networks. However, deploying LLMs demands substantial computational resources, while explicit multi-step channel estimation incurs inherent communication overhead.

To further streamline the optimization process, implicit sensing paradigms entirely bypass explicit channel reconstruction. For MIMO-FAS beam alignment, the authors in \cite{jiang2025beam} propose a channel estimation-free active sensing framework. This approach employs a sequential-activation ping-pong protocol to reconstruct channel observations, circumventing the massive overhead of full-CSI acquisition. Leveraging online and offline learning strategies, it implicitly extracts channel features via pilot interactions, yielding optimal port activation and beamforming to achieve high spectral efficiency without explicit CSI estimation. However, this implicit sensing framework requires iterative online interactions, which may converge slowly in densely connected networks with complex topologies. Furthermore, oriented towards next-generation computing paradigms, a topology-aware quantum GNN (QGNN) is proposed in \cite{recy2026topology} to maximize the sum rate of FAS.  It encodes topological graph features into parameterized quantum circuits using fundamental quantum logic gates and operates unsupervised to optimize transmission configurations. With shallow circuit design and fast convergence, this quantum-classical hybrid machine learning (ML) scheme demonstrates great potential for topology-aware dynamic wireless network optimization. However, QGNN deployment depends on immature quantum processors, and real-time integration into classical mobile hardware is still in its exploratory stages.

%-----------------------------------------------------------------
\subsection{Joint Port Selection and Sensing/Trajectory}
%-----------------------------------------------------------------
% {\bf{\color{red}Here lacks one paragraph to summarize this subsection.}}
FAS provides extra spatial DoF via flexible port selection, while ISAC and UAV systems introduce sensing requirements, platform mobility and highly dynamic wireless environments. Their integration enables joint utilization of antenna reconfiguration, beamforming and trajectory control, but also leads to tightly coupled mixed discrete-continuous optimization problems. AI tools including DRL, recurrent neural networks (RNNs) and attention mechanisms efficiently learn port selection, sensing beamforming and mobility control policies from environmental states. Capturing temporal correlations, inter-node interactions and system dynamics, these methods avoid exhaustive search and improve communication, sensing and positioning performance. 
\subsubsection{ISAC Systems}

% \cite{wang2024fluid,yang2025towards,ghadi2026ai,new2025fluid}
The joint optimization of port selection and sensing beamforming in FAS-aided ISAC systems constitutes a highly non-convex and NP-hard problem, for which AI has emerged as a key enabler \cite{new2025fluid}. Recent studies have investigated both algorithmic innovations for dynamic optimization and the information-theoretic boundaries imposed by finite AI representation capacity.

In algorithmic optimization, DRL has been widely adopted to avoid the intractability of exhaustive search methods. In \cite{wang2024fluid}, an advantage actor-critic (A2C)-based DRL framework integrated with a pointer network is proposed for discrete port selection in multiuser ISAC systems. The framework further incorporates a masked autoencoder (MAE) to extrapolate effectively from partial CSI. Even with only 15\% available CSI, the proposed method achieves promising communication and sensing performance. However, the port selection is restricted to a finite set of candidates rather than to continuous spatial DoF, and the framework primarily focuses on single-target sensing.
To overcome these limitations in multi-target scenarios, the authors in \cite{yang2025towards} present a block coordinate descent (BCD) framework integrated with the DDPG algorithm for continuous antenna positioning. The proposed approach directly generates spatial displacement adjustments for FAS and employs a shaped reward function to balance communication rates, sensing beampattern gains, and movement penalties, thereby ensuring smooth trajectories while avoiding exponential computational complexity. Nevertheless, these methods rely on idealized neural network learning and overlook the performance limitations imposed by the finite capacity of AI representations. 

To bridge this gap, \cite{ghadi2026ai} establishes fundamental information-theoretic bounds by modeling finite AI representation bottlenecks as inherent system constraints. Specifically, the capacity-distortion region is derived from a Gaussian representation model and implemented using a variational information bottleneck-based encoder. This work characterizes the AI bottleneck as additive noise and demonstrates that extending the FAS size enables the system to approach the AI-limited performance bound.

\subsubsection{UAV Systems}
% \cite{feng2025joint,xu2025transformer,fan2025fluid,xu2025fluid,
% xu2025joint,wu2026fluid,wang2024ai}
The integration of FAS into UAV networks leverages both the mobility of UAVs and the position reconfigurability of FAS to enhance spatial diversity and mitigate interference \cite {fan2025fluid}. However, the joint optimization of 3D UAV trajectories and high-dimensional FAS port selection yields a highly coupled, non-convex problem, rendering conventional optimization methods computationally intractable \cite{wang2024ai,wu2026fluid}. To address these challenges, DRL offers remarkable advantages by capturing complex data correlations, relaxing strict convexity requirements, and enabling robust decision-making in dynamic environments.

In \cite{feng2025joint}, the authors investigate a UAV-enabled MEC system and jointly optimize UAV trajectories and FAS port configurations to minimize the long-term average AoI. By adopting a DDPG algorithm with a multi-component reward design, the scheme effectively guides continuous spatial exploration, achieving a 69.6\% performance gain over conventional fixed single-port antenna systems. However, the framework is primarily tailored to static scenarios, limiting its robustness in highly dynamic environments.
To address unpredictable network disruptions caused by UAV malfunctions, \cite{xu2025joint} proposes a resilient framework in which surviving UAVs adjust their 3D trajectories and antenna ports to restore network service. An RL approach based on attention mechanisms and gated recurrent units (GRUs) is developed, where GRUs capture historical state-action sequences and Transformers model temporal dependencies. By aggregating temporal features, the proposed method achieves up to 10\% higher sum rates than value-decomposition RL and non-FAS schemes. Nevertheless, it lacks explicit modeling of dynamic inter-agent dependencies and joint sensing capabilities.
Furthermore, a cooperative 3D UAV positioning framework is introduced in \cite{xu2025fluid,xu2025transformer}, in which one active UAV and four passive FAS-enabled UAVs collaboratively track a moving target UAV by jointly optimizing trajectories and antenna ports to minimize positioning errors. The framework adopts an attention-based recurrent MARL scheme, in which RNNs capture historical state-action information of each agent, while a centralized Transformer models inter-agent dependencies for non-linear global Q-value aggregation. The proposed scheme significantly reduces the average mean squared error (MSE) for positioning compared with both the value-decomposition MARL and Q-learning baselines. 

%-----------------------------------------------------------------
\subsection{Joint Port Selection and Resource Allocation/Offloading}
%-----------------------------------------------------------------

% \cite{ho2025proximal}\\
% \cite{ho2025energy}\\

% \cite{li2025fluid}\\
Integrating FAS with FL and MEC-enabled edge intelligence offers dynamic spatial reconfigurability, privacy preservation, and distributed computing capabilities.
However, the joint optimization of port selection and resource management poses complex non-convex problems, for which DRL offers an effective solution by learning adaptive policies through continuous interaction with the environment. 

To minimize completion latency in FL-aided FAS over multi-carrier NOMA networks, \cite{ho2025proximal} jointly optimizes antenna port selection, CPU frequency, and transmit power. The proposed PPO-based framework enables joint decision-making for discrete port selection and continuous resource allocation. This latency-aware DRL scheme reduces completion time by up to 71.6\% while ensuring fairness in user resource allocation. However, this approach assumes fixed user-subchannel matching and neglects the effects of imperfect CSI in practical environments.
Considering edge power constraints, \cite{ho2025energy} minimizes total energy consumption by jointly optimizing antenna position, transmit power, and local FL iterations. Specifically, Gumbel-Softmax is incorporated into the DDPG framework to generate probability distributions for discrete port and iteration selection. The proposed energy-efficient algorithm achieves average energy savings of 6.5\%-23.2\% while maintaining FL accuracy. Nevertheless, it suffers from dimensionality explosion in large-scale multi-node scenarios and does not fully address the challenges posed by non-independent and identically distributed (non-iid) data.
Extending to FAS-assisted MEC task offloading, \cite{li2025fluid} jointly optimizes antenna port selection, receive beamforming, user transmit power, and computing resource allocation to minimize the maximum system latency. By integrating non-cooperative game pricing into a multi-agent DDPG framework, the proposed method transforms high-dimensional, multi-user power control into a one-dimensional pricing selection problem. Distributed cooperation between user and BS agents enables decision-making within 3 ms, reducing execution latency by 19.1\%-65.8\%. However, frequent state interactions between users and BSs may introduce additional communication overhead in the multi-agent framework.

\subsection{Lessons Learned} 
In this Section, the reviewed studies show that AI has become a core enabler for exploiting the spatial reconfigurability of FASs in user admission, beamforming, channel estimation, sensing, UAV trajectory optimization, and resource allocation/offloading. Its main roles are to reduce port-search and CSI-acquisition overhead, solve highly coupled mixed discrete-continuous optimization, and enable adaptive decision-making under dynamic channels and network conditions. Transformers, RNNs and LLMs capture spatial-temporal correlations; DRL and MARL support port switching and distributed coordination; generative models enable port-state inference and full-channel extrapolation; GNNs and deep unfolding construct scalable learning frameworks. However, practical deployment is restricted by simplified channel models, perfect CSI assumptions, large offline datasets, and computationally intensive architectures. 

%=====================================================================
\section{AI for Pinching Antenna Systems}
\label{sec:ai_for_PASS}
% {\bf{{\color{red}We need a paragraph here to summarize this section, just like the first paragraph of Section~III.}}}

This section examines AI-assisted antenna design in PASS. By enabling multiple antennas' geometric deployments, PASS offers flexible network configuration options to meet the demands of seamless communication coverage. Despite these outstanding features, problems arising from wireless PASS remain challenging due to coupled antenna design, near-field environmental influences, dynamic antenna configurations, and so on, which directly hinder or even bottleneck the use of conventional methods in beamforming design, resource management, and integrated communication and sensing functions. Therefore, recent efforts have paid close attention to using AI techniques to push this boundary. The relevant papers are summarized in Table \ref{tab:learning_pass_summary}.

\begin{table*}[t]
\centering
\caption{Summary of learning-driven joint antenna position optimization in PASS systems.}
\label{tab:learning_pass_summary}
\footnotesize
\renewcommand{\arraystretch}{1.5}
\resizebox{\textwidth}{!}{%
\begin{tabular}{|l|l|l|l|l|l|}
\hline
\rowcolor{myblue}
\textbf{Category} &
\textbf{Ref.} &
\textbf{Learning Paradigm} &
\textbf{Optimized Variables} &
\textbf{Key Technical Idea} &
\textbf{Main Advantages}
\\ \hline

\multirow{8}{*}{\makecell[l]{Joint Position\\ \& Beamforming}}
& \cite{kang2025campass}
& CNN-based DL
& \makecell[l]{PA positions,\\ Tx/Rx beamforming}
& \makecell[l]{Dual-stream residual CNN trained with\\ an unsupervised rate-maximization\\ objective and spacing constraints}
& \makecell[l]{Near-optimal capacity\\ with low online complexity}
\\ \cline{2-6}

& \cite{guo2025graph}
& GNN
& \makecell[l]{PA positions, pinching\\ beamforming, transmit\\ beamforming}
& \makecell[l]{Permutation-equivariant staged GNN\\ that sequentially learns pinching\\ and transmit beamforming}
& \makecell[l]{Scalable across system sizes\\ and enables real-time inference}
\\ \cline{2-6}

& \cite{xiao2025channel}
& \makecell[l]{MoE/\\ Transformer}
& CSI estimation
& \makecell[l]{Position-aware MoE and\\ self-attention architectures supporting\\ arbitrary PA configurations}
& \makecell[l]{Low pilot overhead and\\ strong zero-shot generalization}
\\ \cline{2-6}

& \cite{lv2026deep}
& \makecell[l]{CNN + Attention\\ + BiGRU}
& \makecell[l]{Time-varying\\ CSI estimation}
& \makecell[l]{Cross-attention-based spatio-temporal\\ modeling of antenna geometry\\ and channel evolution}
& \makecell[l]{Robust estimation under\\ Doppler and fast fading}
\\ \cline{2-6}

& \cite{xu2025beamforming,xu2026joint}
& \makecell[l]{KKT-guided\\ Transformer}
& \makecell[l]{PA positions, transmit\\ beamforming, pinching\\ beamforming}
& \makecell[l]{Learns dual variables derived from\\ WMMSE/KKT reformulation instead of\\ complete beamforming solutions}
& \makecell[l]{Improved interpretability,\\ scalability, and performance}
\\ \cline{2-6}

& \cite{jiang2026pinching}
& DRL
& \makecell[l]{PA positions, beamforming,\\ artificial noise}
& \makecell[l]{Mobility-aware RL for joint\\ covert communication and sensing}
& \makecell[l]{Higher covert rates\\ and sensing performance}
\\ \hline

\multirow{7}{*}{\makecell[l]{Joint Position\\ \& Resource\\ Allocation}}
& \cite{xie2025graph}
& \makecell[l]{Graph Attention\\ Network}
& \makecell[l]{Antenna positions,\\ power allocation}
& \makecell[l]{Bipartite graph representation with\\ constraint-aware output mapping}
& \makecell[l]{Scalable unsupervised optimization\\ and millisecond-level inference}
\\ \cline{2-6}

& \cite{karagiannidis2025deep}
& Supervised GNN
& Antenna activation
& \makecell[l]{Attention-enhanced graph learning for\\ binary antenna activation decisions}
& \makecell[l]{Robust under\\ position uncertainty}
\\ \cline{2-6}

& \cite{pakravan2026ai}
& DDPG
& \makecell[l]{PA positions, power allocation,\\ time-switching ratio}
& \makecell[l]{Continuous-control RL for energy-\\ efficiency maximization under battery\\ and location uncertainty}
& \makecell[l]{Significant energy-efficiency\\ improvement}
\\ \cline{2-6}

& \cite{ma2026joint}
& SAC
& \makecell[l]{PA positions, transmit\\ powers, time allocation}
& \makecell[l]{KKT-assisted action-space reduction\\ combined with entropy-regularized RL}
& \makecell[l]{Stable convergence and\\ near-optimal performance}
\\ \cline{2-6}

& \cite{pariyal2026ts}
& \makecell[l]{Hybrid RL +\\ Optimization}
& \makecell[l]{PA positions,\\ NOMA power allocation}
& \makecell[l]{RL-based positioning with\\ analytical power-allocation optimization}
& \makecell[l]{Better sample efficiency\\ and convergence}
\\ \cline{2-6}

& \cite{wu2026straggler}
& \makecell[l]{Fuzzy Logic\\ + DDPG}
& \makecell[l]{PA positions, powers,\\ CPU frequencies}
& \makecell[l]{Hybrid orchestration combining\\ fuzzy client selection and\\ RL resource control}
& \makecell[l]{Reduced FL latency and\\ straggler effects}
\\ \hline

\multirow{4}{*}{\makecell[l]{Joint Position\\ \& Sensing}}
& \cite{gao2026integrated}
& A2C
& \makecell[l]{Segment selection,\\ PA positions, beamforming}
& \makecell[l]{Hysteresis-controlled RL for\\ SWAN-enabled ISAC optimization}
& \makecell[l]{Improved convergence and\\ communication-sensing tradeoff}
\\ \cline{2-6}

& \cite{qin2025joint}
& \makecell[l]{Maximum-\\ Entropy RL}
& \makecell[l]{PA positions,\\ user transmit powers}
& \makecell[l]{Entropy-regularized policy learning\\ for robust ISAC optimization}
& \makecell[l]{Better exploration and\\ reward stability}
\\ \cline{2-6}

& \cite{gao2026rl}
& GNN + A2C
& \makecell[l]{3D PA positions, scheduling,\\ power allocation}
& \makecell[l]{Heterogeneous GNN encoder\\ for topology-aware RL}
& \makecell[l]{Exploits geometric relationships\\ and improves convergence}
\\ \cline{2-6}

& \cite{gao2026llm}
& \makecell[l]{Graph-Enhanced\\ LLM}
& \makecell[l]{Deployment, segment\\ partitioning, beamforming}
& \makecell[l]{CSI-induced graph encoding with\\ LoRA-adapted LLM backbone}
& \makecell[l]{Strong transferability across\\ varying user and target configurations}
\\ \hline

\end{tabular}
}
\vspace{-2em}
\end{table*}

%  {\color{red}(Hoc and Luong(5 pages) )  }
%=====================================================================
% Overview paragraph introducing the section and the 3 subsections
% \hl{I have completed the discussion. Will soon add one figure to summary all methods for sections}
%-----------------------------------------------------------------
\subsection{Joint Antenna Position and Beamforming Optimization}

Unlike conventional arrays with fixed geometries, PASS jointly optimizes PA positions and beamforming, introducing additional spatial DoF but also stronger nonconvex coupling since the channel varies with the array configuration. Although iterative methods such as BCD and WMMSE can address this problem, their complexity becomes prohibitive in large-scale dynamic scenarios. Consequently, AI-based approaches have emerged as an efficient alternative for predicting near-optimal PA configurations and beamforming solutions with low online latency. 
Inspired by this fact, recent efforts on learning-driven PASS focus on the following directions: (i) fast inference via data-driven models, (ii) structure-guided learning that embeds optimality conditions, and (iii) sequential decision making under mobility, sensing, and security constraints. 

\subsubsection{Fast inference via data-driven models}
In \cite{kang2025campass}, a DL framework for capacity maximization, namely Channel-aware multi-pinch antenna selection network (CaMPASS-Net), is proposed for realistic IoT environments. This framework outlines a dual-stream design, in which a residual CNN architecture is designed to jointly optimize transmit precoding and antenna positioning on both the transmitter and receiver sides. Meanwhile, an unsupervised learning strategy with a custom rate-maximization loss is developed to handle complex-valued objectives and antenna-spacing constraints. In this way, CaMPASS-Net can achieve near-optimal capacity performance with significantly reduced online complexity for high-dimensional PASS optimization. In \cite{guo2025graph}, a GNN-based learning framework is introduced to jointly optimize PA positions and transmit beamforming to maximize spectral efficiency by exploiting permutation equivariance between users, waveguides, and PAs. In this framework, a staged GNN architecture first optimizes pinching beamforming and subsequently constructs transmit beamforming with low inference complexity. This approach helps capture both path loss and phase variations induced by PAs, generalize across system sizes, and significantly reduce computational latency compared to optimization-based methods, thereby enabling real-time PASS beamforming. 

\begin{figure}
    \centering
    \includegraphics[width=\linewidth]{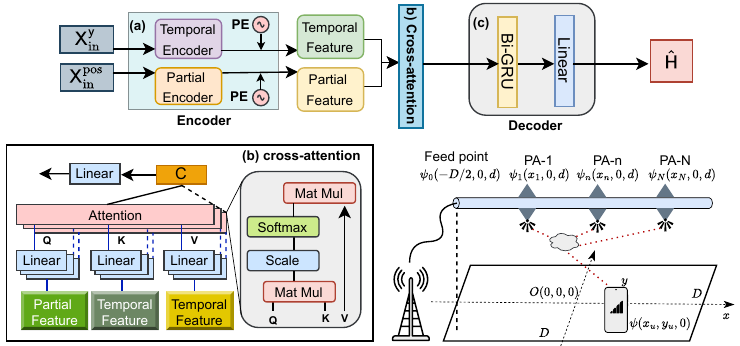}
    \caption{Illustration of a DL-driven spatio-temporal estimator: a) encoder block, b) cross-attention mechanism, and c) decoder block.}
    \label{fig:06}
    \vspace{-2em}
\end{figure}

On another front, two DL-based estimators are proposed in \cite{xiao2025channel} to address the fundamentally ill-conditioned and underdetermined estimation problem caused by multiple PAs sharing a single RF chain per waveguide. In particular, the first estimator, namely PA-MoE, leverages position-aware MoE with dynamic padding and feature fusion to exploit PA location diversity. Meanwhile, the second estimator, namely PAformer, uses a Transformer-based architecture to natively support arbitrary and dynamic numbers of PAs via self-attention. Both designs significantly outperform conventional least-squares (LS)/least-mean-square-error (LMSE) benchmarks under reduced pilot overhead and exhibit strong zero-shot generalization across unseen PA configurations, establishing DL-enabled channel estimation as a key enabler for practical PASS beamforming and resource optimization. Meanwhile, \cite{lv2026deep} proposes a spatio-temporal estimator driven by DL (see Fig.~\ref{fig:06}) to address the problem of time-varying channel estimation when taking into account Doppler effects, multipath fading, and dynamically reconfigurable antenna sets. To cope with the variable number and positions of PAs, PA locations are encoded using a binary activation vector to unify input-output dimensions and preserve spatial information. A heterogeneous encoder then extracts temporal features \({\bf K}\) from pilot signals \({\sf X}_{\rm in}^y \) using multi-scale CNNs and spatial features \({\bf Q}\) from PA positions \({\sf X}_{\rm in}^{\rm pos} \) using MLPs, followed by a cross-attention mechanism that models dynamic correlations \({\bf V}\) between spatial antenna configurations and temporal channel evolution \({\bf K}\). A bi-directional GRU decoder then exploits the spatio-temporal memory of the channel to reconstruct CSI \(\hat{\bf H}\) under fast fading. Simulation results demonstrate that the proposed architecture significantly outperforms LS/minimum MSE and existing DL ratio baselines across varying signal-to-noise ratio (SNR), pilot density, and antenna count settings, while achieving microsecond-level inference, thus establishing an effective and scalable solution for time-varying CSI acquisition in practical PASS deployments.

\subsubsection{Structure-guided learning} 
In \cite{xu2025beamforming}, the authors propose a KKT-guided dual learning (KDL) framework for joint transmit and pinching beamforming in multi-user MISO PASS by reformulating the sum-rate maximization problem using the WMMSE method. This reformulation enables the derivation of closed-form KKT structures for the optimal transmit beamformer, allowing a learning model to predict only a small set of dual variables and PA positions instead of the full set of primal variables. A KDL-Transformer is then developed by modeling CSI-to-beamforming as a sequence-to-sequence task and capturing inter-PA, inter-user, and CSI-beamforming dependencies through attention mechanisms. Numerical results show that this approach outperforms majorization-minimization, penalty dual decomposition, and black-box learning baselines. As an advanced version of \cite{xu2025beamforming}, the authors in \cite{xu2026joint} further develop two solution paradigms: (i) an optimization-based majorization-minimization and penalty dual decomposition algorithm with provable convergence to stationary points, leveraging the WMMSE reformulation and Lipschitz-gradient surrogates; and (ii) a KKT-guided dual learning method implemented with a Transformer architecture, which reconstructs KKT-conditioned solutions by learning only the dual variables. Numerical results demonstrate that the proposed KDL-Transformer surpasses both classical optimization and purely data-driven learning, achieving over 20\% sum-rate gains with millisecond-level inference and establishing a new benchmark for scalable PASS beamforming.

\subsubsection{Sequential decision making}
In this direction, a sensing-aided covert communication framework is introduced in \cite{jiang2026pinching}, empowered by PASS and leveraging its large-aperture near-field characteristics to improve communication covertness and adversary sensing performance jointly. The key idea behind this framework is to integrate extended Kalman filtering with PASS-based probing, enabling the transmitter to accurately track a mobile warden's position and velocity without adversary cooperation. To do so, a joint optimization of beamforming, artificial noise, and PA positions under strict covertness and sensing constraints is decomposed into two subproblems. In which, the first beamforming and AN subproblems admit low-complexity subspace solutions. Meanwhile, the second subproblem of optimizing the PA position is solved using DRL to exploit temporal correlations in mobility. Simulation results verify that PASS achieves higher covert rates and sensing fidelity than fixed-position MIMO.

%-----------------------------------------------------------------
\subsection{Joint Antenna Position and Resource Allocation}
%-----------------------------------------------------------------

% In traditional communication systems, power allocation is optimized over channels with a fixed large-scale structure, so the optimizer primarily balances interference, fairness, and energy efficiency among users. In PASS, however, antenna positions become additional control variables, meaning the propagation geometry itself can be altered before power is allocated. Thus, position decisions affect not only channel strength and user ordering, but also the feasibility and effectiveness of subsequent resource allocation. This transforms the problem from power control over a fixed topology into joint shaping of the topology and the resources, leading to stronger variable coupling, higher-dimensional decision spaces, and tighter physical constraints such as movement bounds, antenna spacing, energy causality, and task-completion requirements. As such, each change in PA placement modifies the underlying channel structure and, consequently, the optimal resource allocation policy. In this case, conventional optimization methods are applicable in limited scenarios with surging computational cost. This burden becomes particularly severe in dynamic systems with uncertain user locations, time-varying battery states, or latency-critical communication-learning interactions.
Unlike conventional systems that allocate power over a fixed channel topology, PASS can jointly adapt antenna positions and resource allocation. PA placement reshapes the propagation environment, directly affecting channel conditions and the resulting allocation policy. This leads to highly coupled, high-dimensional optimization problems subject to physical constraints such as movement limits, antenna spacing, energy causality, and task requirements. Consequently, conventional optimization methods incur substantial computational overhead, especially in dynamic scenarios with uncertain user locations, varying energy states, and latency-sensitive operations.

For this reason, learning-driven approaches have become especially attractive for PASS-based resource allocation. In contrast to iterative solvers that must repeatedly recompute power and scheduling decisions as antenna geometries change, a trained model can infer near-optimal control actions with much lower online complexity. More importantly, the existing literature suggests that PASS-oriented learning methods naturally fall into three levels based on how the decision problem is structured. The first level consists of graph-based or supervised design learning, where the goal is to learn static or quasi-static placement, activation, or power-allocation policies directly from structured input data. The second level consists of continuous-control deep RL, where antenna positions, power variables, and time-sharing parameters are jointly optimized in dynamic environments through interaction with the system. The third level consists of hybrid intelligent orchestration, in which learning is combined with additional reasoning layers, such as fuzzy logic, task-aware classification, or analytical subproblem solvers, to reduce action complexity and better align the learned policy with communication-system objectives. 

\begin{figure}
    \centering
    \includegraphics[width=.7\linewidth]{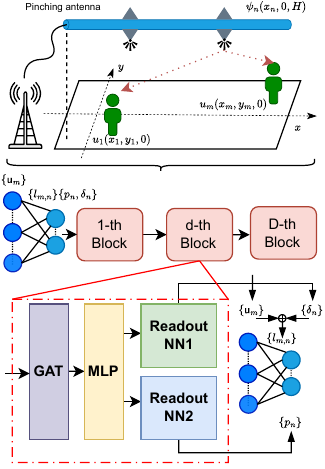}
    \caption{ Graph representation of the considered system and the architecture
of the proposed model in \cite{xie2025graph}.}
    \label{fig:07}
    \vspace{-2em}
\end{figure}
\subsubsection{Graph-based and supervised learning for static or quasi-static antenna design}
In \cite{xie2025graph}, the authors formulate the downlink PASS energy-efficiency maximization problem as a joint antenna placement and power allocation task as a bipartite graph network (BGAT) with two node types, namely user and antenna nodes, connected through geometry-dependent edge features, as illustrated in Fig.~\ref{fig:07}. In terms of design, BGAT combines three functional components in each block: a graph attention module for relational feature extraction, an MLP for feature transformation, and readout neural networks that map latent features into feasible antenna intervals and power values. Besides, to ensure feasible solutions for BGAT when establishing the energy-efficiency objective as the loss, the readout stage is leveraged to rescale interval variables to satisfy spacing and position bounds and to rescale power variables to satisfy the total power constraint. With this design, BGAT is not only suitable for unsupervised training and scalable across different numbers of users, but also can operate with millisecond-level inference and demonstrates superior performance compared to conventional designs and DL benchmarks. 
In \cite{karagiannidis2025deep}, the authors consider a practical PASS scenario where antenna positions are fixed along the waveguide and only binary activation states are optimized. The problem is formulated as a supervised learning problem, with globally optimal activation labels generated offline using Gurobi. Three architectures are investigated: i) an MLP that processes antenna-wise channel features with global fusion, ii) a GNN+MLP that captures antenna-user interactions through message passing, and iii) a GNN plus distributed policy network (GNN+DisPN) that incorporates attention-based distributed decision making. To improve training, a composite loss combines weighted binary cross-entropy with differentiable SNR-aware penalties and collapse-avoidance regularization, encouraging both label accuracy and signal quality. Results under position uncertainty show that GNN+DisPN achieves the best performance by first learning graph embeddings and then applying a two-stage attention mechanism to derive user-aware representations and antenna importance scores for activation decisions.

\subsubsection{DRL for continuous PASS resource control}
In \cite{pakravan2026ai}, a wirelessly powered uplink NOMA network assisted by pinching-antenna arrays is investigated, with a focus on maximizing energy efficiency by jointly optimizing PA positions, user transmit powers, and the time-switching ratio between energy harvesting and uplink transmission. For this problem, the main challenges stem from nonlinear energy-harvesting behavior and uncertainties in user locations and battery states, which makes the problem highly nonconvex and analytically intractable. To address this, the authors first formulate the decision process as a continuous-control MDP and adopt a DDPG-based DRL framework. Then, the learning agent observes battery levels and user locations as states, outputs continuous actions for power control, PA deployment, and time allocation, and is trained using an energy-efficiency-oriented reward with penalties for feasibility violations. With this design, the proposed DRL method significantly improves energy efficiency over fixed-antenna and static-resource-allocation baselines, achieving gains of at least 0.5 Kbps/Hz. Meanwhile, a latency-critical uplink MEC scenario with two users and a single PA is considered in \cite{ma2026joint}, where a hybrid NOMA offloading strategy is employed to minimize total transmit energy, along with DRL solutions. Facing challenges in optimizing variables of the PA position, user transmit powers, and the duration of an additional time slot for the less urgent user, the authors first use a contradiction-based argument under the KKT conditions to prove that the task-completion constraints are active at the optimum, which simplifies the action space and improves the reliability of learning. Then, they rely on this structural reformulation to build an entropy-regularized SAC algorithm, in which the state includes user positions, antenna position, power variables, and time allocation, while actions are defined as incremental updates rather than absolute values to support smoother exploration. In this setting, the reward is defined as the negative total energy plus a penalty for incomplete task delivery, and the SAC agent is trained with twin critics, stochastic policy updates, replay sampling, and soft target network updates. With this design, the designed DRL framework achieves near-optimal performance under dynamic channel conditions while maintaining stability and generalizability.

In contrast, the work \cite{pariyal2026ts} addresses a two-user downlink NOMA-assisted PASS scenario and proposes a two-stage soft actor-critic (TS-SAC) framework to maximize the system sum rate subject to antenna spacing, QoS, and SIC constraints. In this, the key learning idea is to reduce the reinforcement-learning action space. To be specific, instead of letting DRL jointly optimize both antenna positions and NOMA power allocation, the first stage uses an SAC agent to optimize only the continuous antenna positions, while the second stage uses a constrained nonlinear optimizer to solve the NOMA power allocation subproblem.  This decomposition helps preserve the adaptive capability for the most difficult spatial control variable while reducing the burden on the policy network, thereby improving sample efficiency and learning stability. Since TS-SAC formulates a hybrid DRL plus conventional optimization, it can outperform fully end-to-end DRL baselines such as DDPG, TD3, and standard SAC with corresponding improvements of 23.7\%, 11.7\%, and 9\% in the average sum rate at 20 dBm, while illustrating that in PASS not all variables need to be learned jointly if a judicious decomposition can simplify the control problem.

\subsubsection{Hybrid intelligent orchestration for communication-learning systems}
The authors in \cite {wu2026straggler} introduce PASS in the FL domain and propose a hybrid conventional-and-PA network to mitigate straggler effects in uplink FL. In this system, the authors first classify the clients into conventional, pinching, and discarded categories using a fuzzy-logic-based classification scheme, in which the inputs are the communication quality and data contribution of each client. After classification, the authors formulate a total-latency minimization problem that jointly optimizes PA placement, transmit power, and computation frequency in a NOMA-enabled FL setting. Since the resulting optimization is nonconvex and highly coupled, the authors solve it using a DDPG-based DRL algorithm. In this setting, the MDP is structured so that the state includes client dataset sizes, channel conditions, and system variables; the action includes PA position, uplink powers, and CPU frequencies; and the reward penalizes latency while enforcing energy constraints. Moreover, the learning component does not operate in isolation: it is preceded by a fuzzy-logic front end that performs structured client selection before DRL-based resource optimization begins. In this sense, it achieves a hybrid intelligent pipeline in which symbolic rule-based reasoning (fuzzy logic) is combined with DRL-based continuous control to address FL communication bottlenecks. The proposed hybrid network therefore enhances overall FL performance and outperforms a purely beamforming- and resource-allocation-based framework.

%-----------------------------------------------------------------
\subsection{Joint Antenna Position and Sensing}
Joint antenna positioning and sensing design in PASS extends conventional ISAC by making the sensing aperture reconfigurable. By adjusting antenna positions along meter-scale waveguides, PASS can simultaneously reshape the sensing aperture, target illumination, and communication channels. This introduces strong coupling between geometry, signaling, and sensing performance, yielding additional spatial degrees of freedom but also greater nonconvexity, higher dimensionality, and physical constraints. Although conventional optimization methods remain applicable, their complexity increases substantially because every geometry update changes both the communication and sensing channels, requiring repeated optimization of coupled variables.

In this context, learning-driven approaches are attractive because they can amortize the computational cost of repeated joint sensing-communication optimization, rather than solving a new nonconvex program whenever the target, user, or PASS configuration changes, a trained model can infer near-optimal positioning and control decisions with much lower online latency. More importantly, the current literature suggests that learning-driven PASS sensing methods can be classified into three levels. The first level consists of reinforcement-learning-based sequential ISAC control, where antenna positioning, sensing, and communication variables are optimized through repeated interaction with a dynamic environment. The second level consists of graph-structured learning, in which the geometric and relational interactions among antennas, users, and targets are explicitly encoded to improve topology-aware optimization. The third level consists of transferable large-model-based design, where graph-enhanced foundation models are used to learn reusable deployment and beamforming policies that generalize across changing user and target configurations.

\subsubsection{Geometry-aware joint ISAC design with sequential reinforcement learning}
In \cite{gao2026integrated}, an ISAC architecture is built on segmented waveguide-enabled PA (SWAN) arrays, where the BS jointly optimizes segment selection, PA positions, and transmit beamforming under sensing-quality constraints. Then, a segment hysteresis-based RL framework is developed by formulating the MDP over communication CSI, sensing CSI, prior antenna positions, and prior segment assignments, and an A2C algorithm serves as the learning backbone. Instead of updating segment selections freely at every step, this framework considers a hysteresis gate that probabilistically decides whether a new segment-selection action should replace the previous one. This mechanism is designed to reduce structural non-stationarity and stabilize training, since frequent remapping of antennas across segments can make the environment highly volatile. With this design, the developed framework is made segment-aware and transition-stable through hysteresis, thereby significantly improving convergence behavior compared to conventional RL baselines.

\subsubsection{Graph-structured learning for topology-aware ISAC optimization}
For this level, a more compact PASS-assisted ISAC scenario is investigated in \cite{qin2025joint}, where the main focus is on jointly optimizing PA positions and user transmit powers to maximize the total communication rate while satisfying sensing-SNR and system-energy constraints. Using a maximum-entropy-based RL (MERL) algorithm, the problem is reformulated as a sequential decision-making process in which the state includes PA positions, user positions, target positions, and the energy budget. In contrast, the action includes antenna displacement, user movement-related variation, power allocation, and energy consumption. The central learning idea is to train a stochastic policy by maximizing not only the cumulative reward but also the policy entropy, thereby encouraging a balance between exploration and exploitation and improving robustness to dynamic channel conditions. Architecturally, MERL uses a policy network together with twin critic networks to mitigate overestimation, which places it conceptually close to modern entropy-regularized actor-critic methods. Compared with deterministic DRL baselines such as DDPG, it has been shown that explicitly maximizing entropy leads to better reward stability, higher communication rate, and stronger sensing performance. Meanwhile, \cite{gao2026rl} extends PA-assisted ISAC to a 3D deployment setting, in which three orthogonal waveguides provide full-space spatial diversity, and the optimization variables include 3D antenna positions, time scheduling, and user transmit powers. For this system, the authors outline a heterogeneous GNN-based RL framework. In this, a heterogeneous GNN encoder is first used to transform the mapping graph into a structured latent representation, and the resulting graph embeddings are then fed into an A2C-based actor-critic RL module. Instead of treating the environment as a flat state vector, the GNN-RL framework allows the learning model to explicitly exploit entity types and inter-entity relationships in 3D, as it can generate a structural representation of communication and sensing in a coupled-geometry manner. Therefore, it outperforms baseline methods in both performance and convergence.

\begin{figure}
    \centering
    \includegraphics[width=0.9\linewidth]{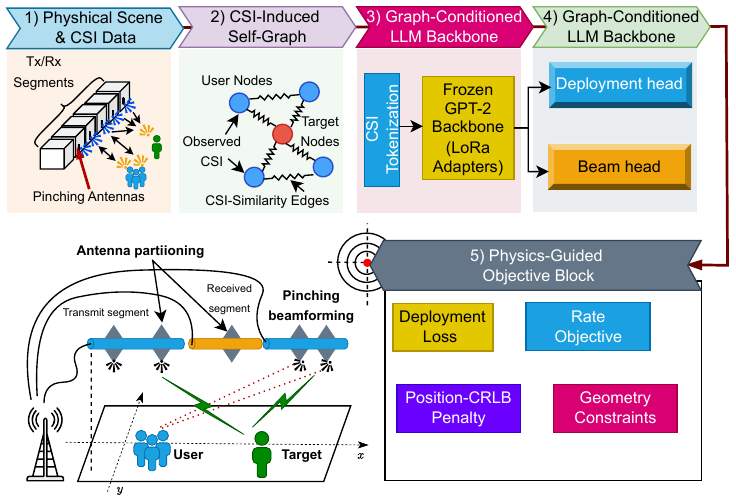}
    \caption{Illustration of the LLM-enabled general learning framework in \cite{gao2026llm}.}
    \vspace{-2em}
    \label{fig:08}
\end{figure}
\subsubsection{LLM-based design for variable user/target configurations}
% {\bf{Transferable foundation-model-based design {\color{red}I feel very weird of this calling, but do not know how to modify it.}}} 
For this level, the authors in \cite{gao2026llm} study a SWAN-ISAC system in which the BS must jointly optimize antenna deployment, segment-wise transmit/receive partitioning, and communication/sensing beamforming under coupled communication-rate and sensing-accuracy requirements, as described in Fig.~\ref{fig:08}. To address this problem, a general learning framework is proposed by combining a CSI-induced self-GNN (SGNN) with an LLM backbone adapted via LoRA. In terms of operation, the SGNN first transforms the CSI of users and targets into a permutation-invariant graph representation, where communication users and sensing targets are treated as graph nodes and their interactions are encoded through CSI-similarity-based adjacency. Then, the resulting graph-enhanced representation is passed into a pretrained GPT-style LLM backbone, which serves as a high-capacity sequence model to capture the nonlinear coupling among deployment, segment partitioning, and beamforming variables. On top of this shared backbone, the framework employs two task-specific heads: one for deployment and segment-partition prediction, and another for beamforming prediction. In this way, the driven-learning framework can capture all system parameters via a graph-enhanced LLM for transferable SWAN-ISAC optimization, where structured graph encoding ensures permutation invariance and LoRA-equipped LLM inference enables scalable, parameter-efficient adaptation across varying communication-user configurations.

%-----------------------------------------------------------------
\subsection{Lessons Learned}
%-----------------------------------------------------------------

Learning-driven PASS studies suggest that intelligence is most effective when combined with analytical and structural knowledge. Early data-driven methods directly map channel or geometric information to PA positions, beamforming vectors, or CSI, substantially reducing online computation but remaining sensitive to distribution shifts \cite{kang2025campass,guo2025graph,xiao2025channel,lv2026deep}. Structure-aware approaches improve stability, interpretability, and scalability by embedding KKT conditions, dual-variable representations, graph relations, attention mechanisms, segmented-waveguide constraints, or permutation-invariant architectures into learning models \cite{xu2025beamforming,xu2026joint,xie2025graph,karagiannidis2025deep,gao2026integrated,gao2026rl,gao2026llm}. For dynamic scenarios, continuous-control DRL methods such as DDPG and SAC can jointly optimize antenna positions, power, and time allocation, particularly when supported by decomposition or analytical reformulation \cite{pakravan2026ai,ma2026joint,pariyal2026ts}. Recent studies further extend PASS learning to closed-loop tracking, sensing, security, FL orchestration, and adaptive ISAC deployment, indicating that PASS should be viewed not only as a beamforming architecture but also as an intelligent system-level controller \cite{jiang2026pinching,wu2026straggler}.

%================================================================
\section{Conclusions, Open Issues, and Future Research Directions}
\label{sec:conclusions}
%================================================================
In this paper, we have provided a comprehensive review of AI for spatially reconfigurable antenna systems, covering MAs, FASs, and PASSs within a unified framework. We first introduced the fundamentals of these three architectures and explained why AI techniques are needed to address the complex design and optimization problems arising in such systems. We then reviewed and compared AI-enabled approaches for each architecture according to the problems they address, including beamforming, channel estimation, ISAC, trajectory and phase-shift control, physical-layer security, and resource allocation. This survey can therefore serve as a useful reference for researchers and engineers by identifying suitable AI techniques for different problems and antenna architectures. Building on these findings, we summarize the key challenges and promising future research directions below.

\subsection{Existing Challenges}
%=======================

\subsubsection{CSI Acquisition Overhead}
Although AI-based methods can reduce CSI requirements, they still rely on
channel observations and therefore cannot eliminate acquisition overhead. Many
existing designs assume perfect or accurate CSI, as exemplified by the
STAR-RIS-assisted MA SWIPT framework in~\cite{amiri2025movable} and the
DL-based MA downlink design in~\cite{kim2025deep}. Robust formulations have
begun to account for imperfect CSI~\cite{xiu2026robust}, but channel
uncertainty is still neglected in many learning-based antenna configuration
schemes. Data-driven extrapolation methods, such as
AGMAE~\cite{zhang2024learning} and SSNet~\cite{gao2025ssnet}, can reconstruct
full-port CSI from sparse observations, with evaluated observation ratios as
low as $5\%$--$10\%$. Nevertheless, their offline training may still require
extensive channel measurements. Moreover, CSI estimation and antenna
configuration are generally treated separately. Jointly learning which
positions or ports to probe and how to configure the antenna system under CSI
uncertainty therefore remains an important research challenge.

\subsubsection{Scalability to Large-Scale and Cell-Free Deployments}
Most existing studies consider relatively small systems with only a few users and a limited number of antenna elements, ports, or PAs, whereas practical deployments may involve much larger network dimensions. The resulting scalability challenges vary across AI models. Value-based DRL becomes increasingly difficult as the discrete action space expands, as observed in pixel-based MA control~\cite{huang2025deep} and discretized movable-element optimization~\cite{geng2025aerial}. Attention-based estimators incur rapidly increasing computational costs as the number of ports grows, motivating linear-complexity architectures such as~\cite{zhao2026taaformer}. Similarly, the message-passing overhead of GNNs increases with the size of the user--port or user--PA graph~\cite{he2025graph}. In cell-free deployments, multi-agent learning introduces additional training, coordination, and information-exchange overhead, while limited fronthaul capacity restricts the amount of information that can be shared across APs~\cite{li2025deep}. Therefore, scaling MA, FAS, and PASS to cell-free networks with many APs and users will require hierarchical, sparse, decentralized, or mean-field learning designs, which remain at an early stage of development~\cite{gu2025partially}.

\subsubsection{Adaptive Edge AI for Dynamic, Heterogeneous, and Resource-Constrained Environments}
Practical 6G scenarios, such as UAV and vehicle-to-vehicle (V2V) networks, are highly dynamic, heterogeneous, and resource-constrained. Rapid and non-stationary changes in obstacle layouts, mobility patterns, and propagation conditions may require offline-trained models to be frequently fine-tuned for specific environments, which is challenging under the limited computing and energy budgets of edge devices. Two promising directions including wireless context engineering~\cite{zhao2026wireless} and world-model-based reasoning~\cite{wang2025dual} can be used. The former provides models with structured information, such as mobility trends and long-term channel statistics, while the latter learns environmental dynamics to support predictive decision making. Combining these approaches with lightweight on-device adaptation offers a promising path toward reliable MA, FAS, and PASS operation in dynamic and resource-constrained 6G environments.

%\cite{zhao2026wireless}, %\cite{wang2025dual}

\subsubsection{Hardware Implementation}
Most learning-based policies are trained under idealized hardware and actuation
assumptions, whereas practical implementations introduce additional costs and
constraints. Examples include actuator delay and movement energy in MA systems
~\cite{zhou2026joint}, switching latency and cost in FAS systems
~\cite{liu2026switching}, and waveguide attenuation, PA positioning errors,
and spacing constraints in PASS~\cite{xu2026joint}. Electromagnetic mutual
coupling is also often approximated using only a minimum-spacing constraint,
which may not fully represent the actual interaction among antenna elements.
RL is attractive for incorporating hardware effects that are difficult to
model analytically, since movement, switching, and positioning costs can be
included directly in the reward function
~\cite{liu2026switching,ma2026joint}. The computational and energy cost of the AI controller itself must also be
considered. Large backbones, such as a LoRA-tuned LLaMA model for
satellite-FAS~\cite{yang2025fas} and a GPT-style model for PASS-ISAC
~\cite{gao2026llm}, may consume a significant portion of the latency, memory,
and energy budgets that spatial reconfiguration is intended to reduce.
Lightweight models, model compression, and dedicated accelerators
~\cite{xu2025toward} should therefore be regarded as essential design
components. Finally, the limited availability of measured MA, FAS, and PASS
datasets means that most policies are trained and evaluated using simulation
models. Developing hardware testbeds and measured datasets is therefore
necessary to evaluate and reduce the sim-to-real performance gap.

\subsubsection{Generalization of AI Models}
Most existing models are trained for fixed system dimensions and propagation settings, which limits their adaptability to changes in antenna configurations, network sizes, user distributions, and interference conditions~\cite{wang2026learning,he2025graph,guo2025graph,kang2025nmap,tang2025deep}. Their performance may also degrade when models trained under idealized channel assumptions are deployed in environments with blockage, mobility, or NLoS scattering. Permutation-equivariant GNNs~\cite{he2025graph,guo2025graph} and few-shot LLM-based methods~\cite{guo2025llm} provide partial adaptability, but reliable transfer across system scales and propagation environments remains largely unexplored. Promising solutions include domain randomization, transfer learning, meta-learning, and pretraining across diverse geometries and channel conditions. In particular, foundation models pretrained on large and diverse datasets could provide reusable representations that can be adapted to specific antenna architectures and deployment environments through few-shot learning or parameter-efficient fine-tuning.

\subsection{Future Research Directions}
%=======================================

\subsubsection{Waveform-Aware AI for Cell-Free Spatially Reconfigurable Antenna Systems}
The joint design of advanced waveforms \cite{sui2021approximate,yi2026error,yi2026non} and spatially reconfigurable distributed antenna architectures \cite{sui2024ris,sui2025performance} remains largely underexplored. In particular, AFDM is well suited to doubly selective and high-mobility channels \cite{sui2025generalized,sui2026mimo}, while its recent extensions further enhance waveform flexibility and transmission performance \cite{luo2026towards}. Meanwhile, integrating MA, FAS, and PASS into CF deployments may enable user-centric cooperation and provide additional distributed spatial degrees of freedom \cite{li2025deep,xiu2026robust}. Future AI-driven frameworks could jointly optimize waveform parameters, antenna configurations, AP cooperation strategies, and user-centric clustering according to channel and mobility conditions. Such integrated designs are expected to improve channel acquisition, cooperative beamforming, interference management, and sensing performance in dynamic CF networks.

\subsubsection{Near-Field and Hybrid-Field Learning}
As antenna apertures increase through meter-scale PASS waveguides, 6DMA
surfaces, and XL-MIMO arrays, the planar-wave assumption becomes inaccurate,
and both angular and distance information must be considered
~\cite{shao2025hybrid,kang2025nmap,yang2026effective}. Addressing this regime
requires more than retraining conventional AI models with near-field channel
samples, since spherical-wave propagation provides physical structure that
generic architectures may not exploit. Promising directions include
spherical-geometry positional encodings, models that support hybrid-field
propagation across different portions of a large array
~\cite{shao2025hybrid}, and physics-informed learning that incorporates the
near-field coupling among antenna positions, propagation phases, and
beamforming variables.

\subsubsection{Integration with Other 6G Enablers}
% Spatially reconfigurable antennas can complement other important 6G
% technologies, including RIS, O-RAN, and non-terrestrial networks (NTNs).
% Existing MA-RIS~\cite{zhuang2025multi}, STAR-RIS-MA
% ~\cite{yu2025learning,amiri2025movable}, and RIS-FAS
% ~\cite{chen2025unified} studies indicate that jointly optimizing antenna
% configurations and surface phase shifts can provide additional spatial design
% flexibility. However, the resulting high-dimensional and distributed
% optimization problems may require hierarchical or multi-agent learning. NTNs
% are another promising application, as illustrated by the use of LLM-based
% channel prediction for satellite FAS with OTFS~\cite{yang2025fas}. In
% addition, O-RAN may provide a practical platform for deploying learned
% reconfiguration policies, provided that standardized interfaces are developed
% to expose antenna positions, ports, and PA configurations as controllable
% network resources.
Spatially reconfigurable antennas are poised to become a key enabler of 6G networks through their synergy with emerging technologies such as RIS, O-RAN, and non-terrestrial networks (NTNs). Recent advances in MA-RIS \cite{zhuang2025multi}, STAR-RIS-MA~\cite{yu2025learning,amiri2025movable}, and RIS-FAS~\cite{chen2025unified} architectures demonstrate that jointly optimizing antenna positions and surface phase configurations can unlock additional spatial degrees of freedom and substantially enhance network performance. However, the resulting large-scale and distributed optimization problems necessitate new AI-native solutions, including hierarchical optimization and multi-agent learning. In NTNs, reconfigurable antennas can provide adaptive beamforming and channel-aware transmission for highly dynamic satellite and aerial links~\cite{yang2025fas}, while AI-assisted channel prediction further improves reconfiguration efficiency. Meanwhile, O-RAN offers a practical framework for deploying intelligent antenna control policies, provided that antenna positions, active ports, and hardware configurations can be exposed through standardized interfaces as programmable network resources. 
% Looking forward, the convergence of reconfigurable antennas, RIS, AI, O-RAN, and NTNs is expected to enable highly adaptive and self-optimizing 6G networks with unprecedented gains in coverage, spectral efficiency, energy efficiency, and reliability.

\subsubsection{Hybrid Spatially Reconfigurable Antenna Architectures}
MA, FAS, and PASS provide complementary forms of spatial
reconfigurability, motivating future architectures that combine selected
features of these technologies. Initial evidence supporting this direction
includes hybrid conventional-and-pinching antenna networks
~\cite{wu2026straggler} and comparative analyses of the spatial multiplexing
capabilities of MA and PASS~\cite{yang2026effective}. Promising research
directions include integrating switchable fluid ports with movable antenna
elements, combining PASS waveguides with reconfigurable feeds or auxiliary
surfaces, and selecting among different antenna architectures according to
the deployment geometry, traffic conditions, and service requirements.
A unified formulation that represents MA positions, FAS port states, and
PA coordinates as architecture-specific spatial control variables could
support common learning and optimization frameworks. Such a formulation may
also facilitate transfer learning across related antenna architectures,
thereby reducing the data and retraining requirements for new deployments.

\subsubsection{Green and Energy-Efficient Systems}
Improving energy efficiency is an important motivation for spatial
reconfiguration, and several existing studies have considered SWIPT and
wirelessly powered systems~\cite{amiri2025movable,ma2026joint}. Future
research should adopt end-to-end energy accounting that includes antenna
movement, port switching, PA positioning, signaling overhead, and the
computational energy consumed by the AI controller. This is necessary to avoid
overestimating the energy savings provided by systems that rely on
computationally intensive models~\cite{yang2025fas,gao2026llm}. Energy
efficiency should also be incorporated directly into the learning process
through energy-aware reward functions, lightweight network architectures,
model compression, and dedicated accelerators~\cite{xu2025toward}. These
considerations are particularly important for IoT devices and handsets with
limited energy and computing resources.

\subsubsection{Security and Privacy}
Spatial reconfiguration can improve physical-layer security by adapting
antenna positions to strengthen legitimate links and suppress eavesdropping
channels~\cite{wang2026moving}. Related security benefits have also been
investigated for FAS~\cite{vega2026exploring} and covert PASS
systems~\cite{jiang2026pinching}. However, learning-based reconfiguration
introduces new attack surfaces. Poisoned training data or adversarial
observations may induce unfavorable antenna configurations, while configuration
decisions may reveal user locations or channel conditions. Federated learning
has been explored in FAS-assisted systems
~\cite{ho2025proximal,ahmadzadeh2025ai}, allowing raw observations to remain
local. Nevertheless, secure aggregation, adversarial robustness, and
privacy-preserving reconfiguration remain largely unexplored.

\subsubsection{Standardization and Practical Deployment}
Practical deployment requires shared experimental and standardization frameworks. Ray-tracing digital twins can provide site-specific training data, while hardware testbeds can capture actuation delays, switching costs, positioning errors, coupling effects, and waveguide losses that are often omitted from simulations. Common benchmarks covering channel models, deployment scenarios, performance metrics, and baseline algorithms are also needed for fair and reproducible comparisons across MA, FAS, and PASS systems. From a standardization perspective, antenna configurations should be exposed as controllable network resources, while hardware capabilities, such as movement range, switching latency, positioning accuracy, and waveguide constraints, should be represented through interoperable interfaces. Together with model-based, offline, and safe RL, these developments can support the transition from simulation studies to practical 6G deployments.

\bibliographystyle{IEEEtran}
\bibliography{REF_MOV_ANTENNA}{}

\end{document}